Université Pierre & Marie Curie - Paris 6

*Synthèse de l'activité scientifique en vue de l'obtention*
*d'une Habilitation à Diriger des Recherches*

**Inna KUPERSTEIN**

U900 Institut Curie/INSERM/Mines ParisTech
Bioinformatics and Computational Systems Biology of Cancer
Institut Curie, rue d'Ulm 26, 75248, Paris

# DECIPHERING CELL SIGNALING REWIRING IN HUMAN DISORDERS

Soutenu le 14 December

Devant le jury composé de :

*Président* :
Guido Kroemer
*Rapporteurs* :
Marie Beurton-Aimar
Lodewyk Wessels
Fabrice André
*Examinateurs* :
Simon Saule
Denis Thieffry
Emmanuel Barillot

# ABSTRACT

In this synthesis I provide an overview on my scientific activity during more than 10 years. The main transversal topic connecting all scientific projects in which I participated, is cell molecular mechanisms implicated in human disorders. Trained as biologist, I have dedicated first part of my career to studying neurodegenerative mechanisms. Whereas, in the second part of my career I focused on deciphering complexity of cancer mechanisms by computational systems biology approaches, which is my current scientific activity, also reflecting the future interest. Four chapters of the synthesis chronologically describe the major areas of my interest:

**Chapter 1** is dedicated to the experimental part of the work on mechanisms of neurodegeneration, with focus on oxidative stress signaling and synaptic toxicity in Alzheimer's disease.

Chapters 2-4 are devotes to the work in the field of systems biology of cancer and represent sequential steps of multidisciplinary project on signaling rewiring in cancer and new therapeutic intervention schemes development:

**Chapter 2** describes how to address complexity of cancer by systematic representation of signaling implicated in the disease in the form of comprehensive signaling network maps and impact of this approach on interpretation of cancer omics data.

**Chapter 3** summarizes studies on application of signaling networks modeling for finding synthetically interacting genes in cancer; predicting drug synergy and suggesting complex therapeutic intervention sets.

**Chapter 4** deals with more fundamental topic of mechanistic principles in synthetic lethality. The first part is dedicated to new synthetic lethal paradigm that we suggested following modeling studies. In the second part, I discuss synthetic lethal mechanisms at several scales of intra- and inter-cellular signaling, from molecules to functional modules in cell and toward synthetic interactions between different cell types.

The synthesis is concluded by describing challenges and future directions in the field of signaling rewiring studies in human disorders, articulating the complementarity of experimental and computational approaches.

# RESUMÉ [FRENCH]

## ETUDE DES RESEAUX DE SIGNALISATION DANS LES MALADIES HUMAINES

Dans ce mémoire je présente un panorama de mon activité scientifique depuis plus de dix ans. Le principal axe transversal reliant tous les projets scientifiques auxquels j'ai participé est celui des mécanismes moléculaires de la cellule impliqués dans les maladies humaines. Biologiste de formation, j'ai dédié la première partie de ma carrière à l'étude des mécanismes neurodegénératifs, tandis que la seconde partie a été consacrée à l'étude de la complexité des mécanismes du cancer par des approches computationnelles de biologie des systèmes. Ces derniers constituent mon activité scientifique actuelle et mon centre d'intérêt futur. Les quatre chapitres de mon mémoire décrivent dans l'ordre chronologique mes domaines majeurs d'intérêt:

**Le chapitre 1** est dédié à la partie experimentale de mon travail sur les mécanismes de neurodégénération, avec un focus sur le stress oxydatif signaling et la toxicité synaptique dans la maladie d'Alzheimer.

Les chapitres 2-4 sont consacrés au travaux dans le domaine de la biologie des systèmes du cancer et représentent les étapes successives d'un projet multidisciplinaire d'étude des réseaux de signalisation du cancer et de la recherche de nouveaux schémas d'intervention thérapeutique:

**Le chapitre 2** décrit d'une part comment appréhender la complexité du cancer par une représentation systématique de la signalisation impliquée dans cette maladie, sous la forme de cartes aussi complètes que possible du réseau de signalisation; il montre d'autre part l'impact de cette approche sur l'interprétation de données omiques de tumeurs.

**Le chapitre 3** résume les études utilisant la modélisation de réseaux de signalisation pour trouver les gènes en interaction synthétique dans le cancer, prédire les synergies entre drogues, et proposer des schémas complexes d'intervention thérapeutique.

**Le chapitre 4** aborde de façon plus fondamentale les principes mécanistiques de l'interaction létale synthétique. La première partie porte sur un nouveau paradigme d'interaction synthtique que nous avons suggéré suite à une étude de modélisation mathématique. Dans la seconde partie, je discute les mécanismes de létalité synthétique à différentes échelles de signalisation intra- et inter-cellulaire, de la molécule aux modules fonctionnels de la cellule, et jusqu'aux interactions synthétiques entre différent types cellulaires.

Le mémoire se conclut par la description des défis et directions futures dans le domaine des réseaux de signalisation des maladies humaines, s'appuyant sur la complémentarité des approches experimentales et computationnelles.

# PREFACE

*Dedicated to the respected and beloved ones*

The knowledge of cell molecular mechanisms implicated in human diseases is expanding and should be converted into guidelines for deciphering pathological cell signaling and suggesting appropriate treatment. The basic assumption is that during a pathological transformation, the cell does not create new signaling mechanisms, but rather it hijacks the existing molecular programs. This affects not only intracellular functions, but also a crosstalk between different cell types resulting in a new, yet pathological status of the system. There is a certain combination of molecular characteristics dictating specific cell signaling states that sustains the pathological disease status. Identifying and manipulating the key molecular players controlling these cell signaling states, and shifting the pathological status toward the desired healthy phenotype, are the major challenge for molecular biology of human diseases.

From the beginning of my career I have been interested in understanding the mechanistic basis of biological functions. Going from the detailed information about cell signaling to an abstract model is the way to find regularities in the functioning of a biological system. I strongly believe that a combination of knowledge about emerging principles of a biological system behavior, together with specific molecular perturbations in each patient will help to come up with individual intervention schemes.

The projects depicted in this synthesis represent a combination of my expertise in experimental sciences, specifically cell signaling, with computational systems biology. Each scientific project inevitably requires development of new appropriated methods. Thus, the description of my work combines together the methodological and the scientific achievements.

My diverse arsenal of knowledge and methods is an asset for multidisciplinary studies represented in this document and also for the ongoing and future to-come projects. Among others, my scientific role is to bring together biologists, clinicians, biostatisticians and developers to match various approaches to the current scientific challenges.

## INTRODUCTION

Neurodegenerative diseases as Alzheimer´s, Parkinson, Huntington´s and cancer are two common chronic disorders in the elderly. Neurodegeneration is characterized by progressive dysfunction and eventual loss of neurons, whereas cancer is associated with uncontrolled and excessive cell proliferation. Therefore, these two types of disease seem to be on opposite ends of the cell growth regulation spectrum. Indeed, inverse association between neurodegeneration and cancer has been observed (Lin *et al.*, 2015, Thinnes, 2012).

The initiation of these diseases has a different nature. Neurodegeneration is associated with increased synaptic and neurotoxicity eventually leading to neuronal death, either caused by, or concomitant with, protein misfolding, aggregation and deposition in the brain tissue (Lim and Yue, 2015). Conversely, in cancer, oncogene activation and accumulation of mutations disrupts cell regulation mechanisms resulting in augmented cell survival and/or proliferation (Wade and Wahl, 2006).

During the disease progression multiple and probably common molecular mechanisms are getting involved, but these processes are regulated in an opposite manner in the two disorders. It is obvious that perturbations of mechanisms involved in cell survival and death regulation are affected differently. Therefore it is important to delineate what are the “switch mechanisms” determining the decision to “die” in the case on neurodegeneration or “repair and live”, in the case of cancer. There might be genetic polymorphisms, or epigenetic mechanisms determining a predisposition to malfunction of an underlying common mechanism, dictating prone-to-death state of cells (‘neurodegenerative phenotype’) or prone-to-survive/grow state of cells (‘cancer phenotype’). Thus, in the ‘neurodegenerative phenotype’, cell may be susceptible to cell death stressors such as aggregated proteins, hyperphosphorylation, oxidation, inflammation or other unknown risk factors. Conversely, cells would have a greater likelihood of surviving under stressors, while concomitantly becoming more susceptible to cancer development if they are bearing the ‘cancer phenotype’ (Behrens *et al.*, 2009). There are several key players in the cell signaling that are discussed in the literature as candidate ‘switchers’ among others, p53, WNt, Pin1, that coordinate between cell cycle to cell death and regulated in cancer and in neurodegeneration in an opposite manner. In addition, it is important to keep in mind that more and more cell functions are discovered to be associated with each one of these disorders, as immune response, angiogenesis, metabolic pathways, etc. (Heppner *et al.*, 2015, La-Beck *et al.*, 2015). These observations are pointing to importance of analysis and comparison of these two disorders at the higher level of signaling rewiring, taking into account the whole complexity of signaling network inside the cell and also investigating the impact of different biological functions, provided by interplay between various cell types.

## *Alzheimer's disease: a disorder of protein misfolding and synaptic failure*

Alzheimer´s disease (AD) is age-related neurodegenerative disorder characterized by accumulation of neurotoxic misfolded and aggregated amyloid-beta peptides (Aβ). The Aβ peptides form pathogenic assemblies ranging from small oligomers to large masses of amyloid and each state of the aggregation is characterized by different toxicity potential. Once believed to be brain region-specific and static, these protein aggregates have been recently shown to propagate their conformation through the brain. The outcome of Aβ accumulation is functional compromise of the nervous system.

It is still open question, which are the critical toxic misfolded assemblies initiating synaptic dysfunctions, whether oligomers or larger aggregates? What are the precise molecular mechanisms initiating dysfunctions of synapses and neural circuits? What is the role of additional biological systems e.g. vascular, immune, in the neurodegeneration? Whether some processes are reversible and whether preventive actions are possible to preclude Aβ accumulation and consequent neurodegeneration?

These questions are mostly addressed experimentally and examples of some discoveries are discussed in the **Chapter 1**. In addition, since there is more and more available omics data accumulated by high-throughput technologies and pieces of information describing various aspects of AD signaling, systems biology approaches gaining importance in the AD research (Xia *et al.*, 2014, Santiago and Potashkin, 2014).

## *Cancer: a systems biology disorder*

Molecular biology of cancer has witnessed two parallel evolutions in the last decades, which are complementary and together contributed to our understanding of the biological mechanisms of the pathology, and to the evolution of its clinical treatment. The first of these evolutions is the careful work of biologists who have deciphered in great detail the intricacy of molecular mechanisms that govern tumor initiation and progression, and reported these discoveries in the form of scientific papers.

The second evolution was made possible by the advent of high-throughput technologies like microarrays and later next-generation sequencing, allowing accumulation of molecular profiles of tumours, with exhaustive description of the mutational landscapes, gene expression patterns and epigenetic modifications for a large number of samples. Nearly, 20 000 tumour samples have been profiled so far by two main international efforts, The Cancer Genome Atlas (TCGA, http://cancergenome.nih.gov) and ICGC (http://icgc.org). These two evolutions have contributed to precision oncology, which can be defined as the use of molecular profiles of tumours and constitutional genome to orient the choice of a therapy for a given patient (Topol, 2014, Berns and Bernards, 2012).

However, despite availability of the omics data there is still no clear understanding of molecular mechanisms that would explain a correlation between the data and particular phenotypes in each case. Taking into account the information about biological signaling machinery in cells may help to better interpret the patterns observed in omics data of tumours. This will allow rationalized medicine approach for patients stratification, drug response prediction and treatment assignment (Barillot *et al.*, 2012, Calzone *et al.*, 2014).

To enable data analysis in the context of molecular mechanisms, the information on these mechanism should be systematically and adequately represented. The knowledge about molecular signaling mechanisms in cells is dispersed in thousands of publications, mostly in human-readable form precluding application of methods and algorithms developed in the field of bioinformatics and systems biology. There is a need in formalized compilation of the knowledge in a computer-readable form. The current solution is representation of relationships between cellular molecules in a form of pathway diagrams found in various pathway databases (Chowdhury and Sarkar, 2015, Bauer-Mehren *et al.*, 2009). As the amount of information about biological mechanisms steadily increases, a different approach for organization, and structuring of this data is essential. The aim is to create more global picture of cell signaling with sufficient granularity of molecular details representation, capturing crosstalks and feedback loops between molecular circuits. For this purpose, comprehensive signaling network maps covering multiple cellular processes simultaneously are more suitable than disconnected pathway diagrams. Our advances and contribution to this field are discussed in the **Chapter 2.1**.

Visualization and analysis of omics data in the context of signaling networks allows better interpretation of the data and verification of deregulated mechanisms (Carter *et al.,* 2013, Krogan *et al.,* 2015). Several developments facilitating visualization and analysis of high-throughput data are shown in the **Chapter 2.2.**

Furthermore, data analysis in the context of signaling networks can help to detect data distribution patterns across molecular mechanisms on the signaling maps, verifying network variables as enriched functional modules ('hot' deregulated areas), key players, 'bottleneck' points (Wang *et al.,* 2015). Correlating those network variables with the phenotype, as drug resistance or patient survival, followed by clustering methods allows to stratify patients according to their integrated network-based molecular portraits and to suggest appropriate intervention scheme (Dorel *et al.,* 2015). Application of signaling network to explain mechanism of drug action in cancer is shown in **Chapter 3.1.**

Structural analysis and modeling of different scenarios (e.g. mutants, fusion, sub-cellular re-localization) using networks allow to verify synthetic interactions between molecular players, explain phenotypes and rise a hypothesis for experimental validation (Cohen *et al.,* 2013). How *in silico* studies of cell signaling can lead to mechanistic predictions, validated in the experimental model is discussed in **Chapter 3.2.**

The extreme case of negative synthetic interaction is synthetic lethality (SL). The classical paradigm explains synthetic lethal interactions as a phenomena where combinations of two gene deletions significantly affects cell viability, whereas single deletion of each one of those genes does not (Kaelin, 2005). Synthetic lethality (SL) provides a conceptual framework for the development of cancer-specific drugs. The idea of SL treatment approach is to take an advantage of the specificities in tumor cells which bearing abnormal function of one of the genes from the synthetic lethal pair. Targeting synthetic lethal partner allows then selective killing of tumor cells, and avoiding or limiting side effects on normal cells (Fang, 2014).

The attempt to identify SL pairs in different cancers can be addressed in several ways. For example, experimental approaches may include classical study of molecular mechanism using knockout cell or animal models. Additional, wider approach is high-throughput screens of synthetic lethality using siRNA, shRNA or CRISPR/Cas9 technologies. The sub-set of this method is gene-drug synthetic lethality screening aiming to retrieve genes-sensitizers for the drug. Those methods already provided a bunch of information of SL gene and gene-drug pairs and lead to generation of SL databases and networks (Measday *et al.*, 2005, Ooi *et al.*, 2006). Several SL targets are assumed as druggble for some cancers (Fece de la Cruz *et al.*, 2014). However, the experimental methods are time and resources consuming and can cover only limited number of SL pairs. Another very significant drawback of these approaches is that they address only pairwise SL interactions, but not bigger SL sets.

According to the current understanding, signaling pathways create a complex network with forward and backward regulatory loops, and many redundant pathways, therefore the synthetic lethality pairs paradigm should be extended to the synthetic lethal sets or combinations paradigm (Garg *et al.*, 2013, Huang *et al.*, 2014). Increasing number of synthetically interacting players above two would expand the experiments due to high number of possible combination that is unachievable. The alternative (or complementary) to the experimental is the computational approach that allows to test *in silico* multiple synthetic interactions combinations considering very big comprehensive signaling networks. Example of this approach application to suggest interventions sets inferred from network analysis with patient data is shown in **Chapter 3.3.**

Data from SL screens and knowledge of signaling networks structure allows to infer the organizational principles of pathways and reduction of pathawys up to abstract models. These models are suitable for mathematical modeling to better understand the system's properties and synthetic relationships between players in the model. Using this approach we performed *in silico* simulations that lead to discovery of new mechanism of SL, as described in **Chapter 4.1.**

In addition, systematic representation of signaling in a form of networks, analysis and modeling collectively can provide enough material to try and retrieve emerging principle of signaling organization and to classify the spectrum of synthetic interactions, in particular synthetic lethality. Classification of synthetic interactions mechanisms at different scales of inter- and intracellular signaling is discussed in **Chapter 4.2**.

Similarly, aforementioned systems biology approaches exploiting comprehensive cell signaling network maps, can be useful for analyzing perturbations in cellular processes not only in cancer, but also in other human disorders, as immune diseases, stroke, cardiovascular diseases and neurodegeneration.

# 1. MOLECULAR MECHANISMS OF NEURODEGENERATION

**CHAPTER AT GLANCE**

The main player associated with Alzheimer's disease is amyloid-beta peptide (Aβ) which is aggregated and deposited in the brain tissues during the course of the disease. The chapter describes studies on the role of metal ions, oxidative stress and lipid composition on the physiological vs. cytotoxic potential of Aβ and regulation of neuronal phosphorylation cascades. Moreover, Aβ peptides vary in their lengths and the balance between different peptides types in the brain dictates their aggregation status. The synaptotoxic and cytotoxic potential of the aggregates and consequent effect on neuronal and cognitive functions are shown. The impact of these discoveries on therapeutic approaches in Alzheimer's disease is discussed.

## *1.1 Phosphorylation cascades and oxidative stress response in neurodegenerative diseases*

*The Weizmann Institute of Science, Rehovot, Israel*

### *Working hypothesis*

During my PhD, I've focused on studying mechanisms involved in oxidative stress response after stroke and consequent neurodegeneration during Alzheimer's disease (AD). It has been known, that the major inducers of neuronal loss in AD are amyloid-beta peptides (Aβ) that gaining neurotoxic properties while their un-controlled accumulation and aggregation. In addition, the AD brains are characterized by increased oxidizes species content. Finally, the correlation between stroke incidents followed by oxidative response and induction of AD symptoms has been found (Zhao and Zhao, 2013). Therefore, the working hypothesis of the project was that oxidation-related processes might be connected to the Aβ-induced neurodegeneration, however the mechanism of action has to be elucidated.

### *Methodologies*

*Primary neuronal cell cultures, mice and rat stroke models were used to study oxidative stress in combination with additional stimuli as Aβ peptides or various pro- and anti-oxidant compounds. I have developed an in vivo method for combination of compound injection with the transient brain ischemia induction in mice and rat models by fine-tuned control of brain blood supply without direct brain surgical intervention. Phosphorylation cascades perturbations in cell models and in the brain tissue from mice and rat models were assessed using basic molecular biology techniques as Western blots, PCR, confocal imaging, antioxidants activity bio-assays, etc*.

***Results***

We have discovered that Aβ peptides, the main players in AD, have dual role in neuronal viability regulation and shown it in primary cell culture and in rat model. Aβ peptides can serve a neuroprotective role by 'buffering' the oxidation agents, but can be switched to neurotoxic compounds during prolonged oxidative stress. (Kuperstein and Yavin, 2003, Kuperstein *et al.*, 2004). In addition, we have found bi-phasic regulation of phosphorylation signaling cascades in neurons involved in the switch mechanism from pro-survival to pro-degenerative (Kuperstein *et al.*, 2001, Kuperstein and Yavin, 2002). The study suggested the new concept of physiological role for Aβ peptide in neurons that has not been considered in the field before. In addition, we've demonstrated one of the possible molecular mechanisms involved in the oxidative stress-dependent neurotoxicity and neuronal death caused by Aβ peptides in their toxic state. These findings were used in development of a combinational treatment scheme in AD with anti-oxidant and anti-aggregation compounds together (Figure 1).

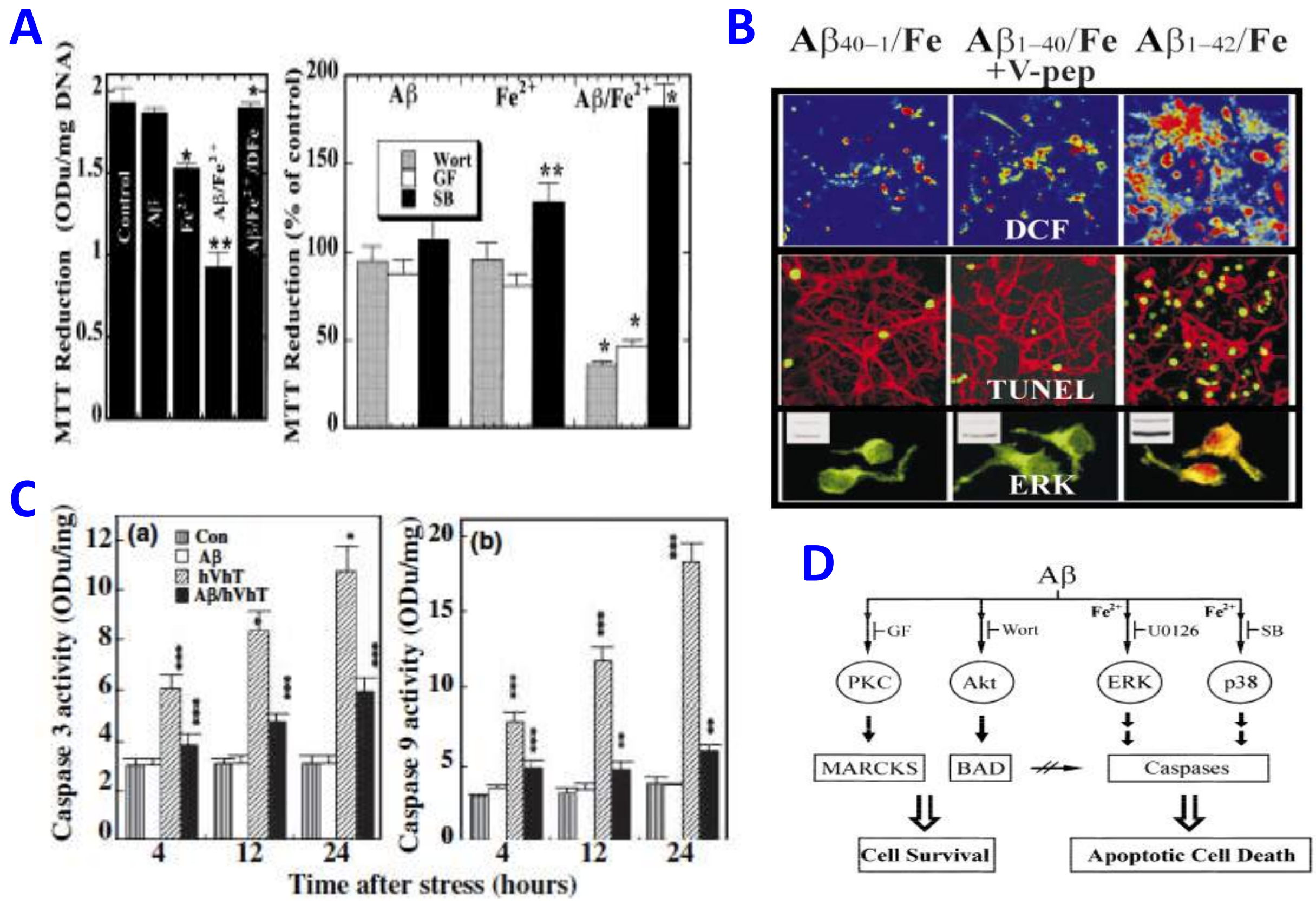

***Figure 1. Dual role of Aβ on neuronal viability****. Effect of different Aβ peptides in presence or absence of metals ions on (A). mitochondrial activity and (B). neuronal toxicity in cell culture, demonstrating that ions are promoting Aβ aggregation (not shown) and consequent neurotoxicity. (C). Physiological concentration of Aβ is neuroprotective against acquit oxidative stress in brain tissue. (D). Kinase inhibitors are used to find out the key kinase pathway involved in the switch between cell survival and cell death.*

***In this collective work I coordinated technician and several students. The project resulted in four major publications, number proceedings publications and collaboration with the pharma start-up company for development of anti-AD treatment schemes.***

## *1.2 Molecular mechanisms of synaptic and neuronal toxicity in Alzheimer's disease.*

*VIB-KU Leuven and IMEC-nanoelectronics and nanotechnology research center, Leuven, Belgium*

### *Working hypothesis*

During my post-doc, I've been involved in two international collaborative projects: (1). MEMOSAD-Verum European network for Memory Loss in Alzheimer's Disease: Underlying Mechanisms and Therapeutic Targets. The project aimed to study amyloid-beta peptides (Aβ) neurotoxic properties and modes of neuronal death in Alzheimer's disease (AD). (2). ASAP-Artificial SynAPse consortium were I participated in design and development of nanoelectronic tool Artificial SynAPse for middle-throughput analysis of synaptic and neuronal toxicity. Artificial SynAPse device has been used during the studies on the neurotoxic potential of Aβ peptides, under the first project.

Aβ peptides are deposited in the amyloid plaques in the brain tissue of AD patients. Therapeutic approaches attempt dissolving those plaques to reduce amyloid levels, considering that the deposits are drivers of the disease. However, a low correlation between the total burden of amyloid deposited in brain and the degree of neurodegeneration in the patients was found lately. AD brains contain heterogeneous mixture of Aβ peptides varying in length, modifications and aggregation potential. The local environment in vicinity of amyloid plaques may affect aggregation status of Aβ peptides (Breydo and Uversky, 2015). The working hypothesis of the study was that Aβ might have non-equal neurotoxic properties and induce various mechanisms of neurotoxicity depending on the mixture composition and aggregation status of the peptides in the brain.

### *Methodologies*

*Biophysical, biochemical and behavior approaches were applied. The mixtures of most abundant Aβ40 and Aβ42 peptides were prepared in various peptides ratios. Combinations of typical natural lipids were added to assess the role of plaques environment in Aβ aggregation. Conformational properties of aggregation species in Aβ mixtures were studied using the transmission electron microscopy (TEM) and Dynamic light scattering (DLS) methods. The aggregation/dissociation kinetics were measured by Fourier transform infrared spectroscopy (FTIR) and ThT fluorescence methods. Primary neuronal cultures were used to assess effect of Aβ mixtures on spontaneous synaptic activity recorded on the Artificial SynAPse device and using the patch-clamping techniques. The composition of synapses was visualized under the*

*confocal microscopy following the fluorescent staining of synaptic markers. The neurotoxicity was assessed using several viability and apoptotic assays. The cognitive and behaviour consequences of brain exposure to the Aβ mixtures were studied in mice models using learning and memory tests as open field-behaviour recording; passive avoidance and contextual/auditory-cue fear conditioning. Finally, distribution of Aβ species in the brain tissue was visualized using the immunofluorescent staining technique.*

### *Results*

**Part I.** It is assumed that Aβ aggregation reaction proceeds 'forward' in an irreversible manner from Aβ peptide to amyloid fibrils found in plaques. We've shown that natural lipids resolubilize amyloid fibrils toward small soluble highly neurotoxic Aβ species that diffuse through the brain and cause memory impairment. The balance between toxic and inert Aβ pools is determined in part by the relative amounts of lipids around the plaques. Therefore, the plaques should be considered as reservoirs of potential toxicity in AD. For example, stroke, trauma or any metabolic perturbations in the brain may result in release of free lipids initiating amyloid fibrils dissociation into toxic Aβ species. The results could also explain why the amount of amyloid deposits and the severity of associated disease symptoms in AD do not necessarily correlate (Martins *et al.*, 2008). In this study we introduced a new understanding of dynamics in the amyloid plaques in AD. The possibility that inert amyloid plaques could be turned into highly neurotoxic species should be considered while designing treatments for AD. Therefore the therapeutic strategies aiming at dissolving amyloid plaques should be critically revised (Figure 2).

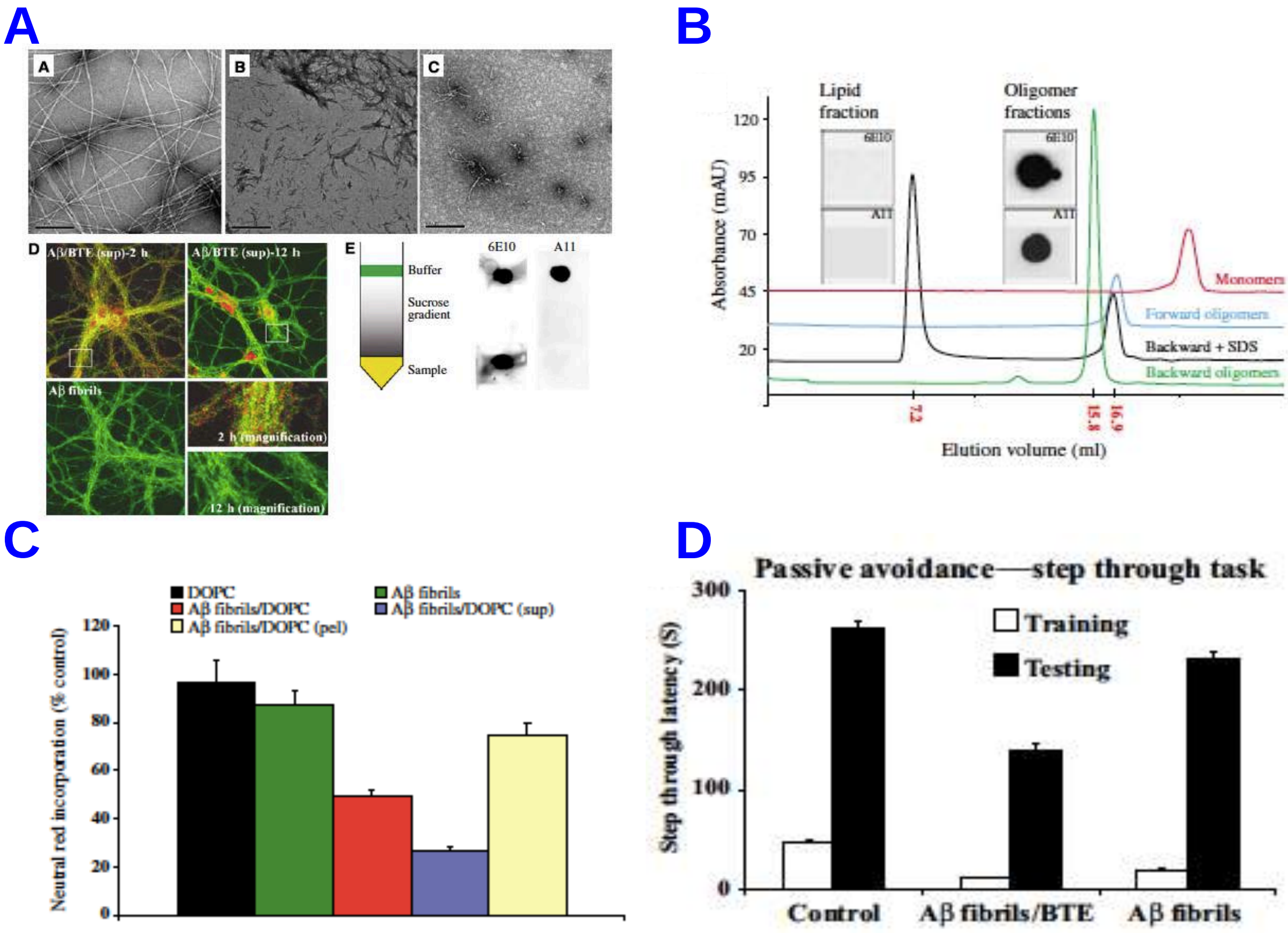

***Figure 2. Lipids induce disassembly of mature Aβ fibrils into soluble toxic oligomers.*** *(A). Electron microscopy micrographs demonstrating lipid-induced dissociation of mature fibrils into oligomes that penetrate to the neuronal cells. (B). The purified fraction of Aβ oligomers (C). causes neuronal cell death and (D). learning and memory formation dysfunction in mice.*

**Part II.** Differences in neurotoxicity and consequent AD severity might be dictated by the equilibrium between the peptides in the Aβ pool. We have shown that a minor increase in the Aβ42:Aβ40 ratio stabilizes toxic Aβ species and activate mechanism of ‘synaptic apoptosis’ involving components of apoptotic machinery that extends lately to the whole cell apoptosis. In addition these Aβ species diffuse in the brain and interfere with learning and memory (Figure 3). The concept that the absolute quantity of Aβ peptides in the brain is less important than the relative Aβ peptides ratio in the pool reflected in their potential to generate stable highly neurotoxic Aβ species was suggested. The finding is important for development of new drugs type aiming to restore the correct ratios between Aβ peptides. This approach promises to lead to the treatment that will prevent neurotoxicity in AD, regardless the amount of Aβ found in the brain (Kuperstein *et al.*, 2010). Following this work, development of new generation of drugs controlling Aβ conformation has been initiated in collaboration with a pharma company.

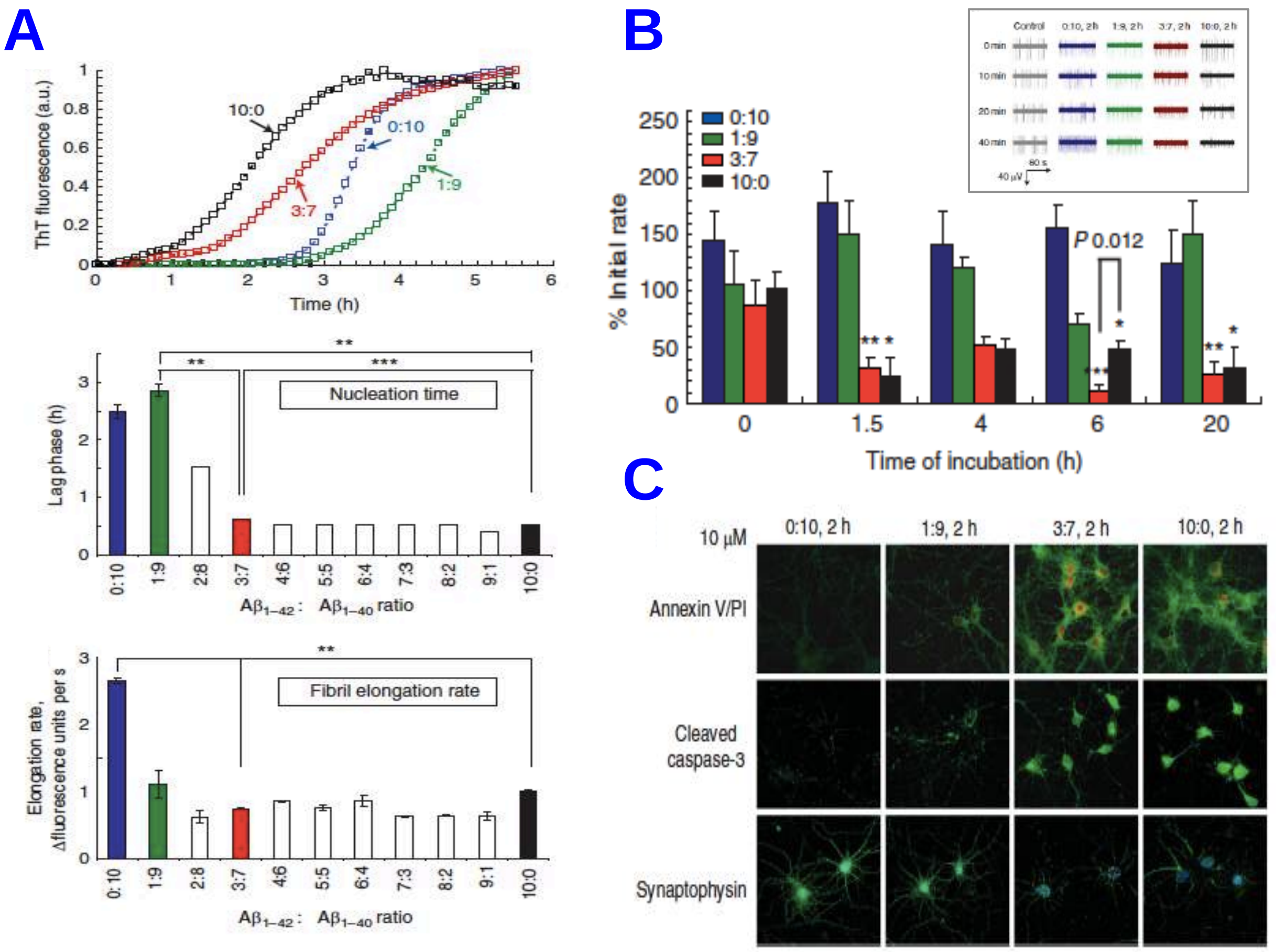

***Figure 3**. **The ratio between Aβ peptides of different length** (A). dictates aggregation dynamics and toxic conformation of oligomers (B). interfering with synaptic firing, (C). synaptic structure and cell viability.*

**Part III.** Detecting effect of neurodegenerative agents at synapses is crucial for considering drugs that would protect synapses at early stages in neurodegenerative process. To allow such developments, synaptic functionalities have to be followed is a systematic way. I have participated in developing a new approach combining neurobiology and nano-scale engineering that lead to construction of Artificial SynAPse device. This neuro-electronic hybrid combines nanoelectronic chip with highly dense and sensitive electrodes; special bio-mimetic surfaces for neuronal growth-on-chip and elaborated signal processing system that allowed to scale down the size of recording area and increase number of recording spots and to improve signal recognition. Currently this and the next-generation Artificial SynAPse devices are used for drugs screening. This project has lead to initiation of the association Neuro-Electronics Research Flanders (NERF) (http://www.nerf.be).

***In these projects I coordinated the collaboration between the teams and supervised technician, students and post-doc involved in the projects. The projects resulted in publication of two major papers, several proceedings; press release on the discoveries; initiation of new drug development; release of nano-electronic device. The work was awarded by the Verum foundation prize.***

# 2. CANCER: A COMPLEX SYSTEM

*Institut Curie, Paris, France*

**CHAPTER AT GLANCE**

The chapter discusses formalization of biological knowledge into a comprehensive map as Atlas of Cancer Signaling Network (ACSN) and Google Maps-based tool NaviCell that supports map navigation. The application of maps for omics data visualization in the context of signaling maps by NaviCell Web Service module is shown. Finally, new tool NaviCom is presented, allowing generation of network-based molecular portraits of cancer using multi-level omics data. The chapter summarizes the achievements of multidisciplinary projects involving numerous collaborations. Ongoing efforts, future plans and perspectives are outlined.

## *2.1 Formalization of biological knowledge as signaling network maps*

### *Working hypothesis*

Deregulation of molecular mechanisms leading to cancer concerns various processes such as cell cycle, cell death, DNA repair and DNA replication, cell motility and adhesion, cell survival mechanisms, mechanisms of immune system angiogenesis and tumor microenvironment.  Usually most of them are involved in the same tumor and modified as the tumor evolves. It is assumed that in pathological situations the normal cell signaling network is affected by deregulated coordination between pathways or disruption of existing molecular pathways, rather than by creating completely new signaling pathways and molecular interactions. The most common abnormalities in pathological situations are perturbations at the level of gene expression, protein abundance or protein posttranslational modifications, irregular 'firing' or silencing of particular signals, wrong sub-cellular localization of particular molecules and so on.

Such quantitative rather than qualitative network changes compared with normal cell signaling could be studied in the context of comprehensive signaling networks by analyzing experimental data obtained from cancer samples, cancer-related cell lines or animal models. This approach helps to understand interplay between molecular mechanisms in cancer, deciphering how gene interactions govern hallmarks of cancer (Hanahan and Weinberg, 2011) in specific context and use this knowledge to stratify patients accordingly. This will lead to new therapeutic strategies, rationalizing the use of targeted inhibitors (Dorel *et al.*, 2015).

An advantage of representing the biological processes in a graphical form is demonstrating collectively multiple cross-talks between components of different cell signaling processes. This allows understanding the global picture and connectivity between processes that is very difficult to keep in mind just from reading multiple scientific papers. Once the processes are depicted together as diagrams, the relationship between molecular circuits in cells can be appreciated, which makes signaling network maps also didactic tools.

For computational systems biology of cancer, this approach dictates the following strategy: 1) represent formally and in sufficient amount of details the existing knowledge about those molecular processes whose involvement in cancer clearly demonstrated; 2) collect and integrate existing quantitative data on cancer genesis and progression and develop methods to analyze them in the light of the knowledge of a normal cell; 3) create mathematical models able to describe distortions of normal cell functioning as a cause of cancer, and to predict the effect of various perturbations; 4) use data analysis and mathematical modeling to suggest new therapeutic schemes.

Despite existence of a large variety of pathway databases and resources (Chowdhury and Sarkar, 2015), only few of them depict processes specifically implicated in cancer and none of those resources depicts the processes with sufficient granularity. In addition, pathway browsing interfaces and data integration tools are not very advanced. The current project aims to formalize knowledge on cancer-related processes as comprehensive signaling network maps, developing algorithms for map navigation, data analysis in the context of maps and data interpretation for basic research and in clinical studies.

***Methodologies and Results***

Construction and update of Atlas of Cancer of Signaling Networks (ACSN) involves manual mining of molecular biology literature and participation of the experts in the corresponding fields (http://acsn.curie.fr). ACSN differs from other databases because it contains more comprehensive description of cancer-related mechanisms retrieved from the most recent literature, following the hallmarks of cancer. Cell signaling mechanisms are depicted using CellDesigner tool (Kitano *et al.*, 2005) at the level of biochemical interactions, forming a large network of 4600 reactions covering 1821 proteins and 564 genes and connecting several major cellular processes. Currently ACSN contains representation of molecular mechanisms that are frequently deregulated in cancer such as cell cycle, DNA repair, cell death, cell survival, and epithelial to mesenchymal transition (EMT). Cell signaling mechanisms are depicted on the maps in great detail, together creating a seamless map of molecular interactions, presented in the form of a global 'geographic-like' molecular map (Figure 4A). ACSN has a hierarchical structure, composed of interconnected maps of biological process implicated in cancer. Each map is further contains functional modules mainly corresponding to canonically-defined signaling pathways (Figure 4C).

The navigation interface include features such as scrolling, zooming, markers and callouts using Google Maps technology adopted by NaviCell (Kuperstein *et al.*, 2013). Semantic zooming in NaviCell (http://navicell.curie.fr), providing several view levels on maps achieved by gradual exclusion of details and abstraction of information upon zooming out (Figure 4B). ACSN is associated with a web-based blog system WordPress for collecting feedback from the community on content of the map that facilities maintenance and updating of ACSN (Kuperstein *et al.*, 2015).

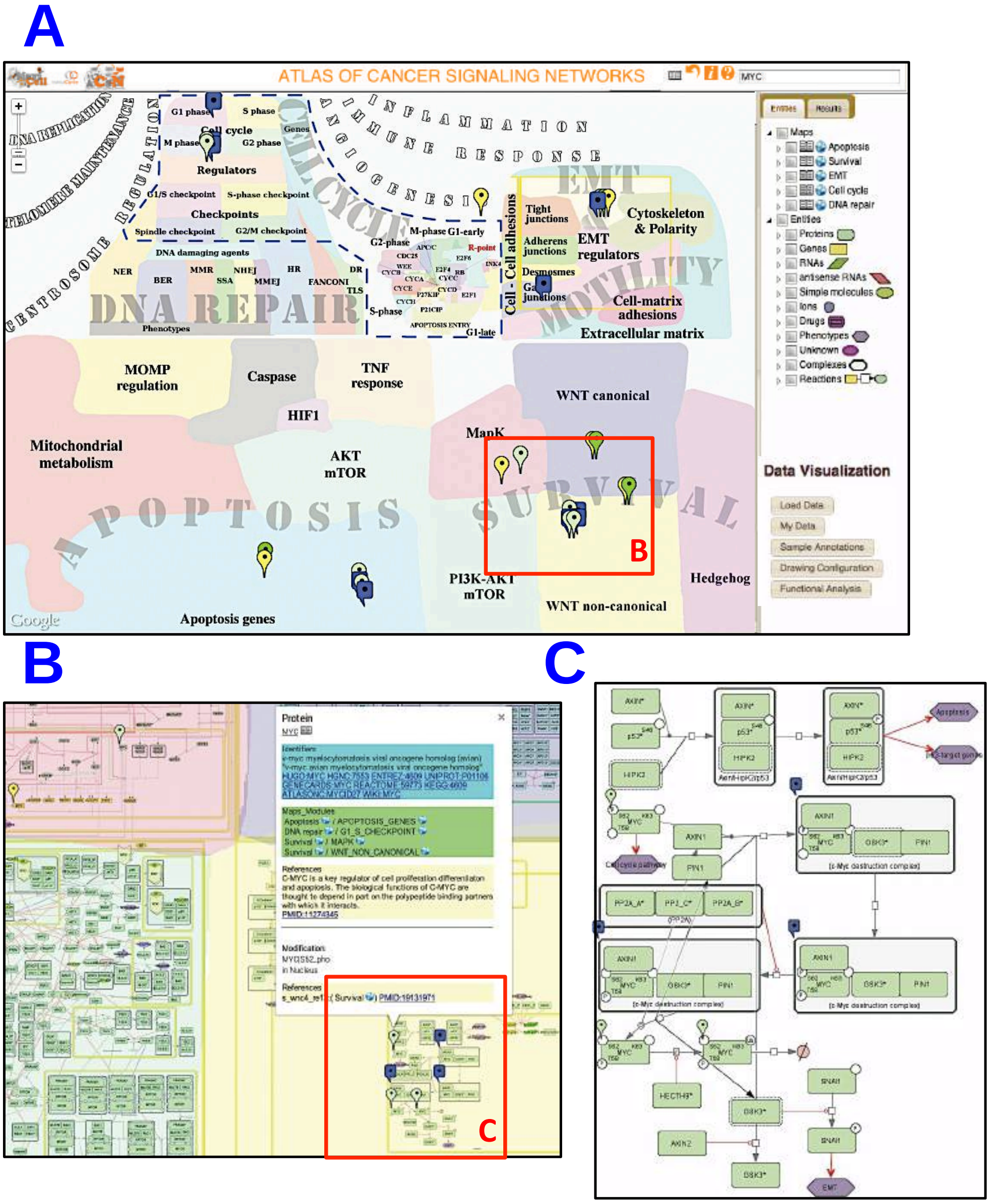

***Figure 4. Atlas of Cancer Signaling Networks.*** *(A). Distribution of frequently mutated oncogenes across human cancers visualized on the ACSN maps; (B). Google-like features of NaviCell for visualization and annotation of map entities; (C). Zoom in on a survival map to observe signaling processes.*

ACSN is a unique resource of signaling in cancer that did not exist in the field, the amount of information embedded and organized in ACSN is enormous. Together with NaviCell, it optimized for integration and visualization of cancer molecular profiles generated by high-throughput techniques, data from drug screenings or synthetic interactions studies. Integration and analysis of these data in the context of ACSN may help in understanding the biological significance of the results, guiding the scientific hypothesis and suggesting potential intervention points for cancer patients. In addition, since ACSN covers major cell signaling processes, the resource and associated methods for data analysis using ACSN are suitable for applications in many biological fields and for studying various human disease.

The atlas is being extended with additional maps depicting molecular mechanisms of DNA replication, telomere maintenance, angiogenesis, immune response and others that will be integrated into future releases of the atlas. The atlas will cover not only intracellular, but also extracellular processes as tumor microenvironment. An additional level of complexity will be added to the atlas in a near future, representing different types of cells surrounding tumor, and interplay between them, to enable modeling of complex phenotypes.

## *2.2 Molecular portraits of cancer: data visualization and analysis using signaling network maps*

The data integration into the ACSN is possible using NaviCell Web Service, an user-friendly environment embedded into the NaviCell tool, allowing to upload several types of "omics" data (expression data for mRNA, microRNA, proteins, mutation profiles, copy-number data) and visualize them simultaneously in the context of molecular interaction maps. Depending on the nature of data, different types of visualization modes can be required to achieve the informative picture (heat maps, bar plots, glyphs and map staining). The data can be visualized at different zoom levels. Sample annotation files unloaded together with the data can serve for defining groups of samples. A novel mode of data visualization for continuous data (e.g. expression) provided by NaviCell Web Service is a 'map staining'. Using the background of the map for visualizing the values mapped to individual molecular entities or group of entities (e.g. score of functional module activities) results in colorful background of the network map that represents the data distribution pattern (Bonnet *et al.*, 2015). All those approaches for data integration into the signaling maps described above allow to rationalize the information embedded into the data: compare samples or group of samples; find typical patterns of data distribution across the molecular mechanisms depicted on the maps; grasp deregulated ‘hot area’ on the maps and major involved players and draw hypothesis as to which mechanisms to concentrate the work in the samples under study. These signaling network-based molecular signatures of samples thus help to stratify patients or samples (Figure 5).

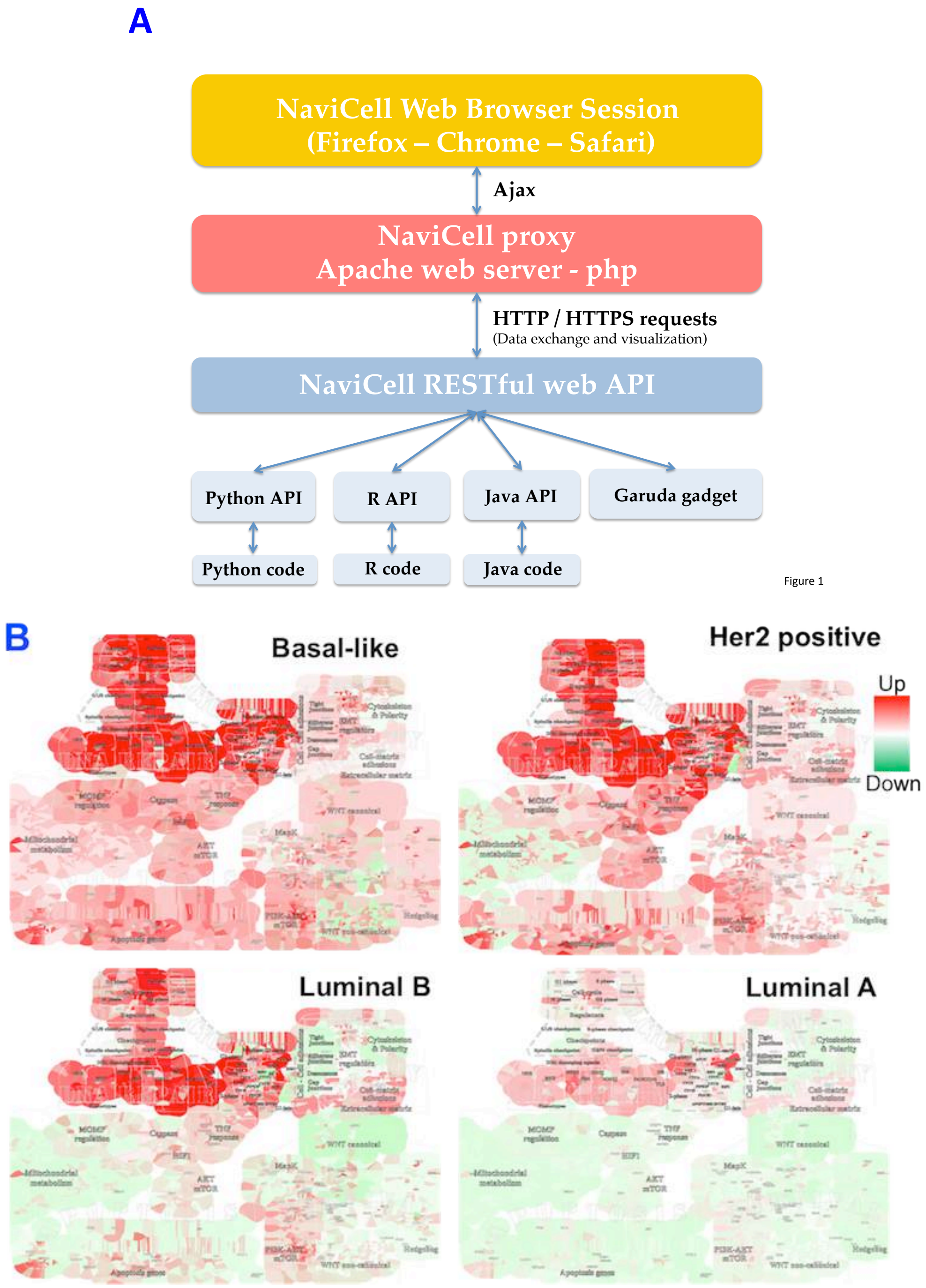

***Figure 5. NaviCell Web service.*** *(A). General architecture of the NaviCell Web service server. Client software (light blue layer) communicates with the server (red layer) through standard HTTP requests using the standard JSON format to encode data (RESTful web service, dark blue layer). A session (with a unique ID) is established between the server and the browser (yellow layer) through Ajax communication channel to visualize the results of the commands send by the software client. (B). BC gene expression data integration and analysis using ACSN. The mRNA expression data from TCGA collection has been used for evaluation of functional modules activities and ACSN coloring as 'map staining' for four BC types. The four BC subtypes are characterized by different patterns of module activities.*

Various omics data are available on the public and local databases (Lapatas *et al.*, 2015). However, there are no tools that support import of big datasets from these databases and displaying them on signaling network maps in efficient way and with optimized visualization settings. To answer to this demand, we developed NaviCom, a python package and web interface for automatic simultaneous display of multi-level data in the context of signaling network map (http://navicom.curie.fr). NaviCom is bridging between cBioPortal database and NaviCell interactive tool for data visualization (http://navicell.curie.fr). NaviCom is empowered by a cBioFetchR R package to import high-throughput data sets from cBioPortal to NaviCell and navicom Python module allowing automatized simultaneous visualization of multi-level omics data on the interactive signaling network maps using NaviCell environment. NavCom proposes several standardized modes of data display on signaling networks maps to address specific biological questions (Dorel *et al*, in revision). (Figure 6A).

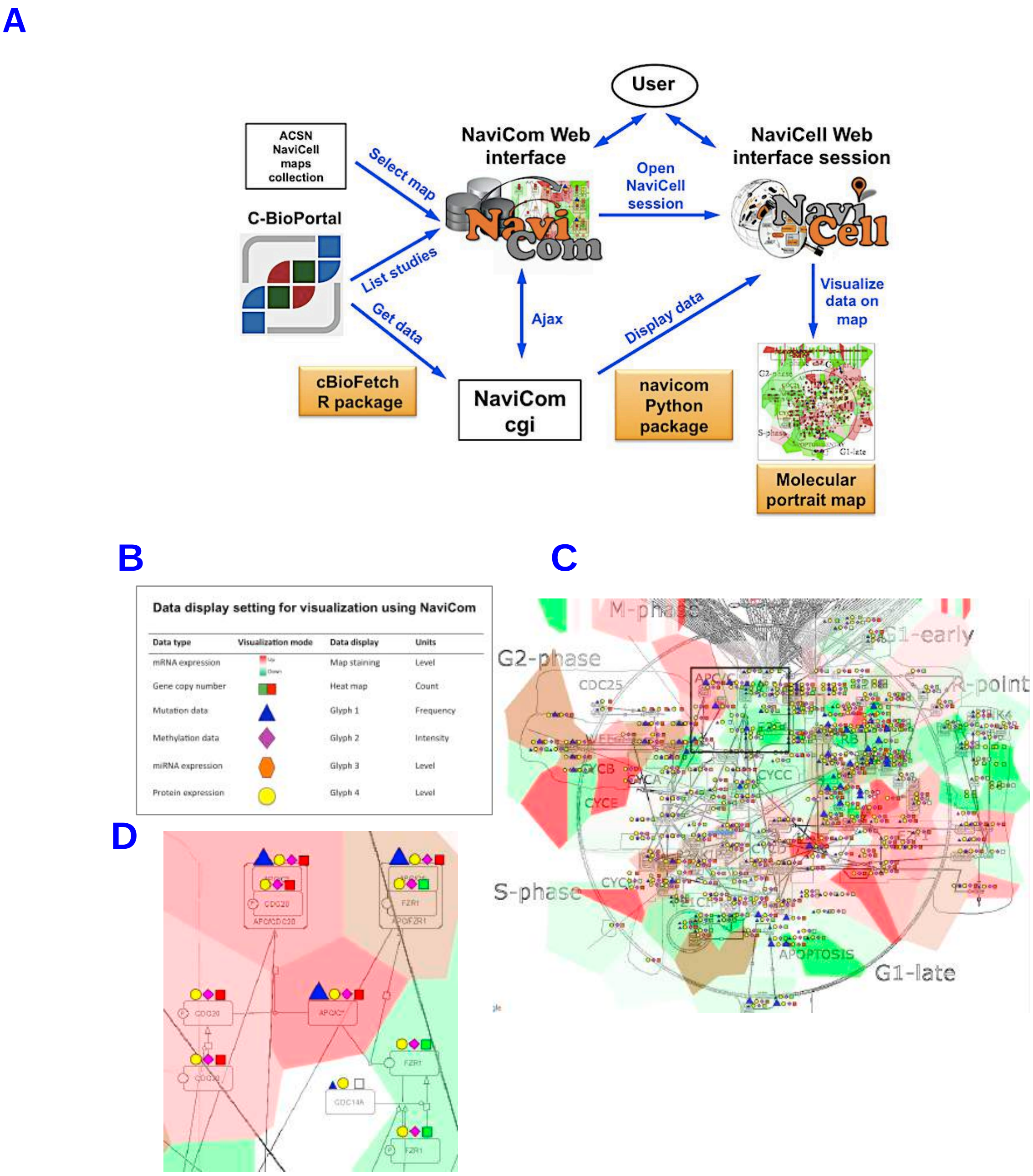

*Figure 6.* ***NaviCom. (A).*** *General architecture of NaviCom environment. The NaviCom interface provides the user with an updated list of studies from cBioPortal and links to ACSN and NaviCell maps collections. When visualization is launched, NaviCom starts a new NaviCell session and calls a cgi on the server. The cgi downloads cBioPortal data to the NaviCell session and displays them to generate the molecular portrait selected by the user. (B). Molecular portrail of Breast Invasive Carcinoma (Nature 2012) from 825 samples visualized on ACSN Cell cycle map. Visualization settings: expression-map staining/ copy number-heat map/ mutations-blue triangle/ methylation-pink diamond/proteomics-yellow circle.*

This tool enables generation of complex molecular portraits from multiple omics datasets from cBioPortal. We aim to create signaling network-based molecular portrait for each disease and studied samples in the context of ACSN or any map prepared in NaviCell format (Figure 6B).

In near future the NaviCom platform will be extended and will provide access to any type of omics data from wide range of databases (TCGA, ICGC, HGMB, METABRIC, CCLE). In addition, to allow broader description of molecular mechanisms implicated in studied sample, signaling networks available in databases as Kegg (Kanehisa *et al.*, 2012), Reactome (Croft *et al.*, 2010) and others, will be also integrated and used for high-throughput data analysis via NaviCom platform.

***I lead this ongoing muntidisciplinary project and supervise the activities of the team. During the projects, five original papers, one book chapter; dozen of proceedings; press releases on the developments have been published. Some of the papers were selected to the highlights talks at international computational biology conferences and were topics for invited seminars. The project has been awarded by the "Thought leader award" grant from Agilent supporting the ongoing collaboration with the Agilent Genespring team for integration of ACSN/NaviCell and GeneSpring features. ACSN, NaviCell and NaviCom are in the process of joining the Garuda Alliance (http://www.garuda-alliance.org), the integrative international platform for systems biology and biomedical research and also the basis for numerous collaborative projects.***

# 3. NETWORK MODELING IN PRE-CLINICAL RESEARCH

*Institut Curie, Paris, France*

**CHAPTER AT GLANCE**

This chapter is dedicated to applications of signaling networks for basic research and pre-clinical studies. In the first project we created differential network-based molecular signatures of sensitivity of two DNA repair interfering drugs. This study allowed us to suggest drugs synergy that has been confirmed for breast cancer cell lines. In the second example, I show how network analysis and modeling help to reveal key regulators of invasion. The prediction allowed to develop the transgenic mice model of early invasive colon cancer. In the third project, we performed a structural analysis of signaling network together with omics data from ovary cancer patients resistant to genotoxic treatment. Following this study we retrieved synthetic lethal gene sets and suggested intervention combinations to restore sensitivity to the treatment. The approaches developed for these projects represent a more general paradigm applicable for other studies.

## 3.1 Explaining synergistic effect of combined treatment in cancer

### *Working hypothesis*

Using DNA repair inhibitors to target cancer cells is a promising therapy but its application is limited by the compensatory activities of different repair pathways. For example, PARP inhibitors that act as synthetic lethal with BRCA deficiency, appear however less efficient in patients with active Homologous Recombination (HR) repair (Lord *et al.*, 2015). During treatment, some tumors escape through compensatory mutations that restore the HR activity or stimulate the activity of alternative repair pathways such as Non-homologous End Joining (NHEJ).

A new class of DNA repair pathways inhibitor (Dbait or DT01) has been recently developed, consisting of 32bp deoxyribonucleotides DNA double helix that mimics double strand breaks (DSB). It acts as an agonist of DNA damage signaling thereby inhibiting DNA repair enzyme recruitment at the damage site (Quanz *et al.*, 2009). However, study of Dbait effects on multiple types of cancer cell lines shows occurrences of resistance in cancer type-independent manner. Since such promising treatments are facing some hurdles including acquired resistance, finding sensitizing agents to restore response to treatment in cancer is needed.

### *Methodologies and Results*

Depending on genetic background, different breast cancer tumors vary in their sensitivity to DNA repair inhibitors, as PARP inhibitors and Dbait. To understand molecular mechanisms underlining these differences, a combination of experimental and bioinformatics approaches was applied. Triple Negative Breast Cancer (TNBC) cell lines were studied for their sensitivity to Dbait (DT01) and the PARP inhibitor Olaparib showing wide distribution of responsiveness to these drugs, that in many cases in not correlated (Figure 7A). We performed integrative analysis of omics data from these cell lines covering mRNA expression, copy number variations and mutational profiles and found that at least 70 non-overlapping genes were robustly correlated with sensitivity to each one of the drugs (not shown). Analysis of the omics data in the context of ACSN maps confirms that different specific defects in DNA repair machinery are associated to Dbait or Olaparib sensitivity (Figure7B). In addition, we identified group of deregulated functional modules across ACSN specifically associated with each one of the drugs and established a predictive drug response network-based molecular portraits. These molecular signatures highlighted different involvement of mechanisms between cells sensitive/resistant to Dbait and Olaparib, suggesting a rational for combination of these two drugs. We confirmed synergistic therapeutic effect of the combined treatment with Dbait and PARP inhibitors in TNBC, while sparing healthy tissue (Figure 7C) (Jday et al., in preparation).

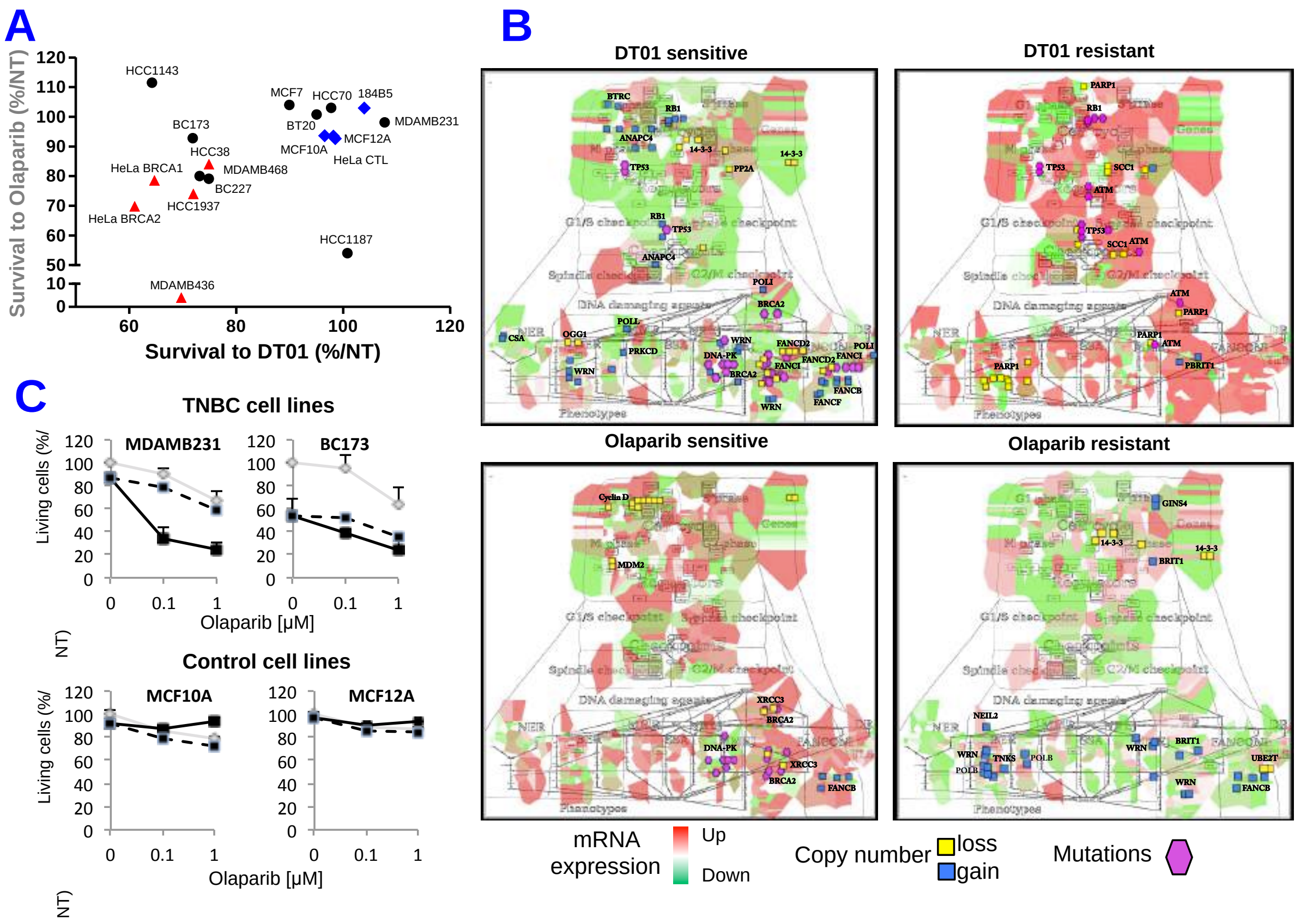

*Figure 7.* ***Sensitivity of TNBC cell lines combination of DNA repair inhibitors.*** *(A). Correlation analysis of survival to DT01 and Olaparib in TNBC and control cell lines. (B). Molecular portraits of DT01 and Olaparib sensitive/resistant TNBC cell lines visualized on*

*DNA repair map. (C). Cell survival to combination of DT01 and Olaparib. With DT01(black line), without DT01(grey line),dashed lines indicate calculated cell survival for additive effect of two drugs.*

***In this ongoing project I supervise scientific activity of a technician and co-supervise a PhD student.***

## *3.2 Finding metastasis inducers in colon cancer through network analysis*

### *Working hypothesis*

Evolution of invasion and metastasis, in particular in colon cancer, has been studied in experimental models, however, the mechanism that triggers the process is still not clear and the available mice models of colon cancer are far from being satisfactory (Hung *et al.*, 2010, Trobridge *et al.*, 2009). With aim to create an experimental mouse model of invasive colon cancer, one needs to address the question what are the major players and the driver mutations inducing invasion. One of early events of metastasis is assumed to be epithelial to mesenchymal transition (EMT) (Nieto, 2011).

### *Methodologies and Results*

In order to identify interplay between signaling pathways regulating EMT, we manually created a signaling network in CellDesigner tool (Kitano *et al.*, 2005) based on the information retrieved from around 200 publications (Figure 8A). This signaling map is now integrated into the ACSN. We performed structural analysis and simplification of the EMT network that highlighted the following EMT network organization principles, which is in agreement with current EMT understanding: (1) Five EMT transcription factors SNAIL, SLUG, TWIST, ZEB1 and ZEB2 that have partially overlapping sets of downstream target genes can activate the EMT-like program. (2) These key EMT transcription factors are under control of several upstream mechanisms: they are directly induced at the transcriptional level by the activated form of Notch, NICD, but are downregulated at the translational level by several miRNAs (namely, mir200, mir34, mir203 and mir192) that are under transcriptional control of p53 family genes. Interestingly, some key EMT transcription factors can inhibit microRNAs, in this way sustaining their own activation. (3) According to the network structure, all five key EMT transcription factors should be activated ensuring simultaneous activation of EMT-like programme genes and downregulating miRNAs. In addition, the EMT key inducers also inhibit apoptosis and reduce proliferation. (4) The activity of Wnt pathway is stimulated by transcriptional activation of the gene coding for b-catenin protein by Notch-induced TWIST or SNAI1. The Wnt pathway, in turn, can induce the expression of Notch pathway factors, creating a positive feedback loop. In agreement with other studies, the Wnt pathway does not directly induce EMT, but helps to maintain it. (5) Components of the Wnt and Notch pathways are negatively regulated by miRNAs induced by the p53 family (p53, p63 and p73). The balance between the effect of positive (Notch and Wnt) and negative (p53, p63 and p73 mediated by miRNAs) regulatory circuits on EMT inducers dictates the possibility of EMT phenotype (Knouf *et al.*, 2012, Moes *et al.*, 2012, Siemens *et al.*, 2011, Fre *et al.*, 2009).

Based on those features of the network we performed network complexity reduction using BiNoM up to core regulators of EMT, apoptosis and proliferation that were preserved through all levels of reduction (Bonnet *et al.*, 2013). The reduced network has been used for comparison between the wild type and all possible combinations of single and double mutants for achieving EMT-like phenotype (Figure 8B,C).

We predicted that in Apc-/-/p53-/- double mutant, Wnt activation does not induce EMT, because p63/p73 induced microRNAs can still inhibit the Wnt and Notch pathway. When Notch is activated in an Apc-/- background, cell proliferation is increased as has been observed in the mice models. At the same time, EMT is inhibited by the microRNA expression induced by, resulting in non-invasiveness. This explains the observed absence of metastases in NICD/Apc-/- mice. Finally, in NICD/p53-/- double mutant, other members of p53 family cannot rescue the function of p53 anymore as constitutively activated Notch inhibits the activity of both p63 and p73. NICD activates the transcription of the EMT key inducers and inhibits the production of microRNAs by suppressing p63/p73 in the context of p53-/-. Notch and p53 have opposite effects on EMT inducers, and overexpression of NICD and knocking-out p53 should have synergetic effect. Furthermore, EMT inducers may activate the Wnt pathway, possibly resulting in a positive feedback loop that will amplify Notch activation and maintain an EMT-like program. Therefore, our computational analysis of the signaling network leads to the prediction that the simultaneous activation of Notch and loss of p53 can promote an EMT-like phenotype (Figure 8D, E).

To validate this hypothesis, we created a transgenic mouse model expressing a constitutively active Notch1 receptor in a p53- deleted background, specifically in the digestive epithelium. Importantly, green fluorescent protein (GFP) expression linked to the Notch1 receptor activation allows lineage tracing of epithelial tumor cells during cancer progression and invasion (Figure 8F). These mice develop digestive tumours with dissemination of EMT-like epithelial malignant cells to the lymph nodes, liver and peritoneum and generation of distant metastases (Figure 8G). We have explored early inducers of the EMT program in human disease and confirmed in invasive human colon cancer samples that EMT markers are associated with modulation of Notch and p53 gene expression in similar manner as in the mice model (Figure 8H), supporting a synergy between these genes to permit EMT induction (Chanrion *et al.*, 2014).

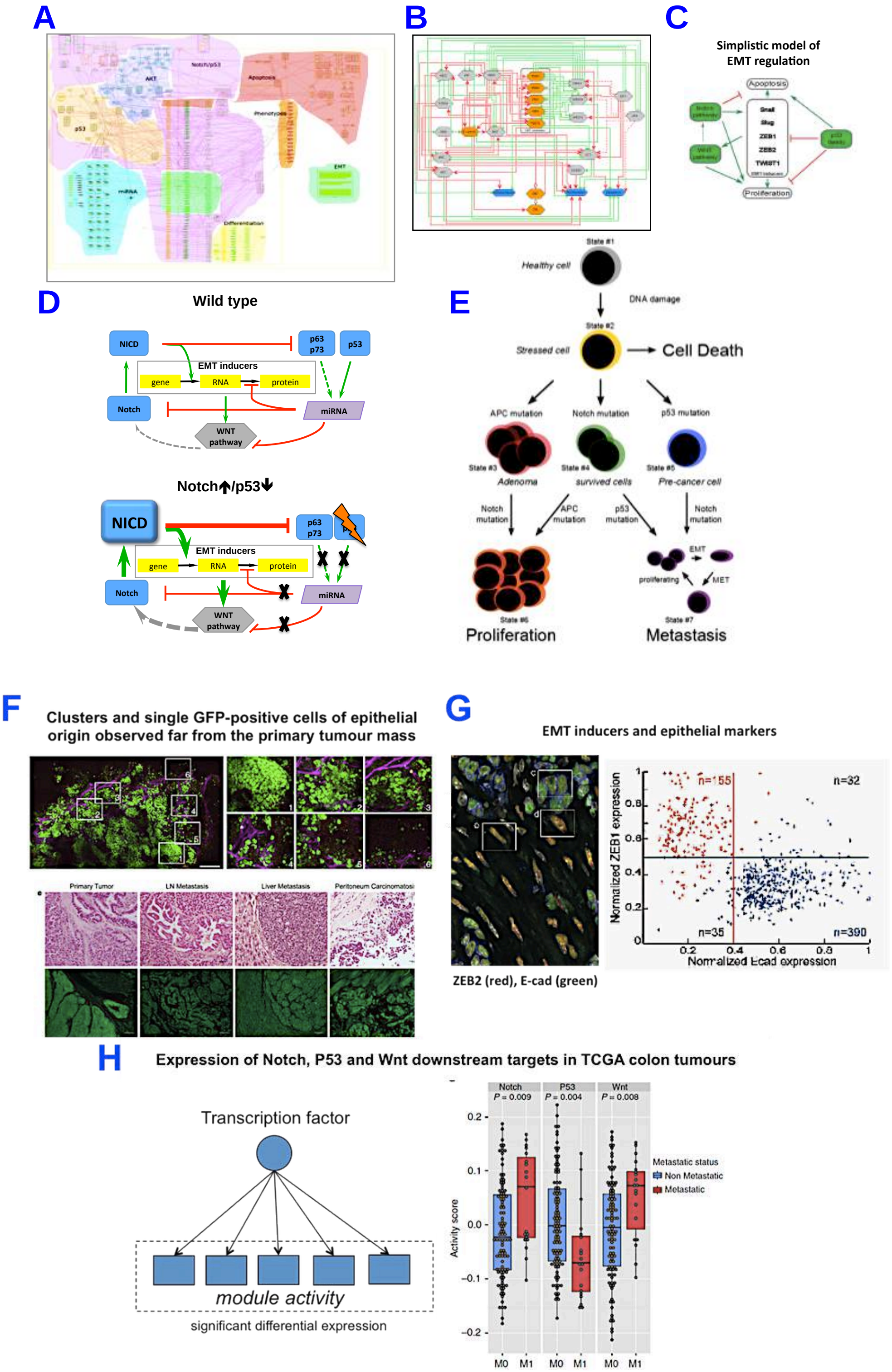

***Figure 8. Prediction of synthetic interaction combination to achieve invasive phenotype in colon cancer mice model*** *(A) Comprehensive signaling network of EMT regulation; (B). Scheme representing major players regulating EMT after structural analysis and reduction of signaling network complexity; (C) and (D) Mechanistic model explaining EMT inducers regulation involving Notch, p53 and Wnt pathways; (E) Phenotypes in single and double mutants; (C) Lineage tracing of cell in tumor and in distant organs; (F) Immunostaining for major EMT markers; (G) Regulation of p53, Notch and Wnt pathways in invasive colon cancer in human (TCGA data).*

Our prediction of synthetic interaction between Notch and p53 demonstrated that there are alternative ways to reach permissive conditions to induce EMT, in addition to those already described in the literature. This idea was not intuitive and actually contradictory to the commonly accepted dogma in the colon cancer field. The study evokes an important message that gathering cell signaling mechanisms together may undercover un-expected interactions and lead to discovery of new mechanisms in regulation of cell phenotypes that may significantly affect the understanding of basic molecular processes implicated in cancer and change the therapeutic approaches. In addition, the comprehensive EMT signaling network is reach resource of information can be used in further studies. Finally, the new EMT mice is a relevant model mimicking the invasive human colon cancer and a system for therapeutic drugs discovery (Kuperstein *et al.*, 2015).

***In this project I supervised scientific activity of one postdoc. The projects resulted in publication of two papers, several proceedings and press communication. Some of those papers were selected to the highlights talks at international computational biology conferences and were topics for invited seminars.***

## 3.3 Complex intervention gene sets derived from data-driven network analysis for cancer patients resistant to genotoxic treatment

### *Working hypothesis*

The idea of SL treatment approach is to take an advantage of the specificities in tumor cells which display abnormal expression or function of one gene from synthetic lethal pair. Targeting synthetic lethal partner allows then selective killing of tumor cells (McLornan *et al.*, 2014). This approach is applied in BRCA2 mutated breast canser cases using PARP inhibitors, however there is frequent escape from the treatment, requiring to more complex solution. One of the reasons for treatment failure is the robustness of cell signaling network ensured by redundant mechanisms that provide the possibility to bypass drugs effect (Dietlein *et al.*, 2014). Therefore the ways for identifying and blocking those active compensatory pathways should be found.

One of the approaches is taking into account the signaling network structure and find the most optimal synthetic lethal combinations of genes (most probably more than pairs) (Huang *et al.*, 2014, Acencio *et al.*, 2013, Zeng *et al.*, 2014). Thus, analyzing synthetic lethal (SL) combinations in the context of patien's omics data as mutation profile, genome, transcriptome, epigenome, etc. to prioritize and chose the appropriate SL set of genes. These analyses may contribute to network-based patients stratification and prediction of sensitivity to traditional drugs, but also help suggesting rationalized treatment schemes adjusted to each patient.

***Methodologies and results***

Genotoxic treatment as Cisplatin, that induces un-repairable DNA damage and cell death specifically in cancer cells, is often inefficient due to backup mechanisms in the DNA repair signaling. To overcome Cisplatin resistance in ovarian cancer, we looked for synthetically interacting combinations of genes to suggest intervention gene sets.

A comprehensive map of cell cycle and DNA repair signaling network constructed from literature curation was used for this study (https://acsn.curie.fr/navicell/maps/dnarepair/master/index.html). The map is composed of three interconnected cell cycle, DNA repair and checkpoints layers covering the most recent knowledge on molecular mechanisms implicated in these processes (Kuperstein *et al.*, 2015). We derived a state transition graph from the map including all paths leading to repaired DNA and genes regulating each step (Figure 9A). Using OCSANA algorithm for searching the minimal cut sets (MCS) on the state transition graphs, considering genes that regulate each step as potential target for interference (Vera-Licona *et al.*, 2013), we identified MCSs whose knock-out completely abort DNA repair (Figure 9B). The coherence of the method has been validated using experimentally–proven SL pairs, verifying the enrichment of the SL pair in real vs. randomly-generated (pseudo)-MCSs (Figure 9C).

For selection of the best MCS to be targeted, we need to find those where part of the components are already altered in the patient, allowing to exploit this background and targeting the remaining components in the set, achieving synthetic lethality. MCSs were evaluated for each patient (TCGA ovarian cancer dataset) using genomic, expression and mutation data integration and correlation with patient's resistance/sensitivity to Cisplatin. The top-correlated MCSs were ranked according for the mutation status of each gene. Those MCSs that were enriched with genes harboring inactivating mutations, were suggested as best intervention combinations to restore sensitivity to Cisplatin by inhibiting the remaining 'active' genes in the set (Figure 9C, table insert). This approach is relevant for complementing genotoxic chemotherapy by targeting specifically cancer cells and exploiting certain defects in the DNA repair machinery in each patient (Russo et al., in preparation).

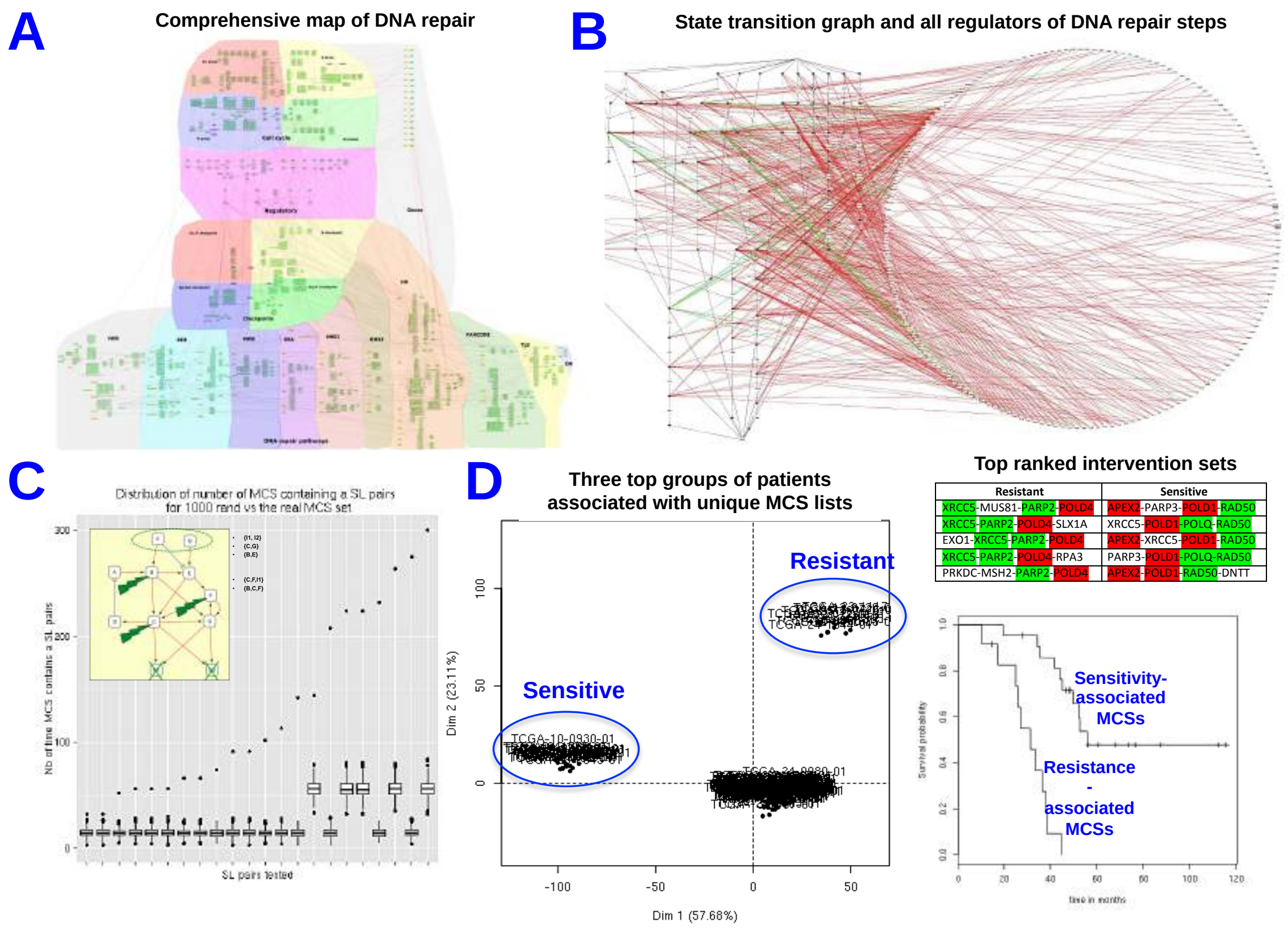

| Resistant | Sensitive |
|---|---|
| XRCC5-MUS81-PARP2-POLD4 | APEX2-PARP3-POLD1-RAD50 |
| XRCC5-PARP2-POLD4-SLX1A | XRCC5-POLD1-POLQ-RAD50 |
| EXO1-XRCC5-PARP2-POLD4 | APEX2-XRCC5-POLD1-RAD50 |
| XRCC5-PARP2-POLD4-RPA3 | PARP3-POLD1-POLQ-RAD50 |
| PRKDC-MSH2-PARP2-POLD4 | APEX2-POLD1-RAD50-DNTT |

***Figure 9. Intervention gene sets for Cisplatin-resistant ovary cancer patients**. (A). Comprehensive map of DNA repair. (B). State transition graph of DNA repair map with regulators of each state transition (regulators of level 1). (C). Enrichment of SL from shRNA screen lines (DECIPHER project) in 'real' vs. pseudo-MCSs. (D). Projection of top 3 principal components of TCGA ovarian cancer patients groups associated with unique MCSs. Survival curves comparing Cisplatin resistant and sensitive patients* associated *with unique MCS sets. Table: Top 5 intervention sets for Cisplatin resistant and sensitive groups (green-inhibited gene, red-activated gene).*

***In this ongoing project I supervise scientific activity of a technician.***

# 4. COMPEX SYNTHETIC INRETACTION MECHANISMS

*Institut Curie, Paris, France*

**CHAPTER AT GLANCE**

This chapter discusses various aspects of synthetic interactions in cell signaling. The first part of the chapter describes how the mathematical modeling of homologous recombination pathway combined with the analysis of synthetic lethal screens lead to discovery of a new type of synthetic lethality mechanism. The second part reviews different models of synthetic lethality mechanisms at several scales, starting from molecular complexes, through molecular pathways and up to the functional modules and different cell types. Taking into consideration this systematic knowledge will help to significantly broaden the canonical interpretation of synthetic interactions, in particular synthetic lethality, with direct implications to molecular pathways understanding and therapy approaches.

## *4.1 Kinetic trap of a pathway: new mechanism of synthetic lethality*

***Working hypothesis***

The classic interpretation of synthetic lethality stipulates that two synthetic lethal genes work in parallel, mutually compensatory pathways (Bandyopadhyay *et al.*, 2010). However, a significant number of synthetic interactions are caused by defects in genes participating in the same molecular pathway (Michaut *et al.*, 2011). The canonical interpretation of pathways assumes only forward propagation. However, the view of molecular pathways as unidirectional, linear reaction cascades is too simplistic. Pathway steps can be reversible which leads to forward and backward propagation of molecular events along the pathway and increases robustness and fidelity of the process.

***Methodologies and results***

The homologous recombination DNA repair pathway is one of the rare examples of a pathway were forward and backward steps are well described (Heyer *et al.*, 2010). In addition, it has been recently shown that one of the intermediates of homologous recombination process is toxic to cells if accumulates above certain concentration for sufficient time period (Figure 10A).

To better understand the system properties of genetic relationships in this pathway, we represented homologous recombination (HR) in a form of simplest model of DNA repair pathway with reversible steps and alternative compensatory pathway. The simplest model contains three statuses of DNA (Substrate-damaged

DNA, Intermediate, Product-repaired DNA) (Figure 10B). We performed analytical study of dynamic features using the simplest linear mathematical model (Figure 10C). We have recapitulated observed genetic scenarios, among others, classical synthetic lethality between pathways (BRCA2-PARP) and the RAD54 mutants that refers to the situation when genetic background dictates susceptibility to accumulation of genetic instability (Figure 10D). We proposed a novel mechanism, within-reversible-pathway synthetic lethality that involves reversible pathway steps, catalyzed by different enzymes in the forward and backward directions. The cell death happens due to kinetic trapping of a pathway into the step resulting in toxic intermediate accumulation because of the mutations in the enzymes catalyzing 'in' and out' reactions (Figure 10D, scenario '4'). The prediction has been validated in Srs2-Rad54 mutants in yeast, where one of the intermediates of homologous recombination process is toxic to cells if accumulates above certain concentration for sufficient time period due to invalidating mutations of the Srs2 and Rad54, catalyzing 'in' and 'out' reactions of the corresponding step.

In order to assess the probability of appearance of within-pathway synthetic lethality in other processes, we ranked all pathways from the KEGG database (Kanehisa *et al.*, 2012) according to their normalized proportion of synthetic lethal interactions within the pathway calculated using yeast gene interaction screen results (Costanzo et al., 2010). Interestingly, HR ranks at the top among DNA repair pathways (Figure 10E).

There is considerable evidence that many molecular pathways include reversible steps catalyzed by different enzymes in the forward and backward directions. Any of those processes can be theoretically trapped into one of their intermediate states if two regulators of forward and backward steps are inactive. In these cases, kinetic trap can be due to the accumulating intermediate or blockade of proper signal propagation or perturbed resource recycling, etc. (Figure 10F).

**A**

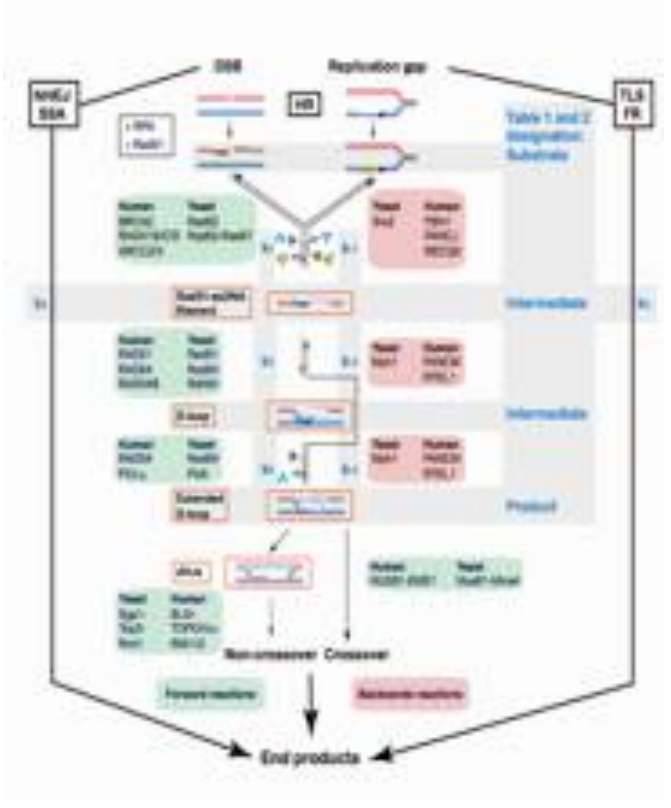

**B**

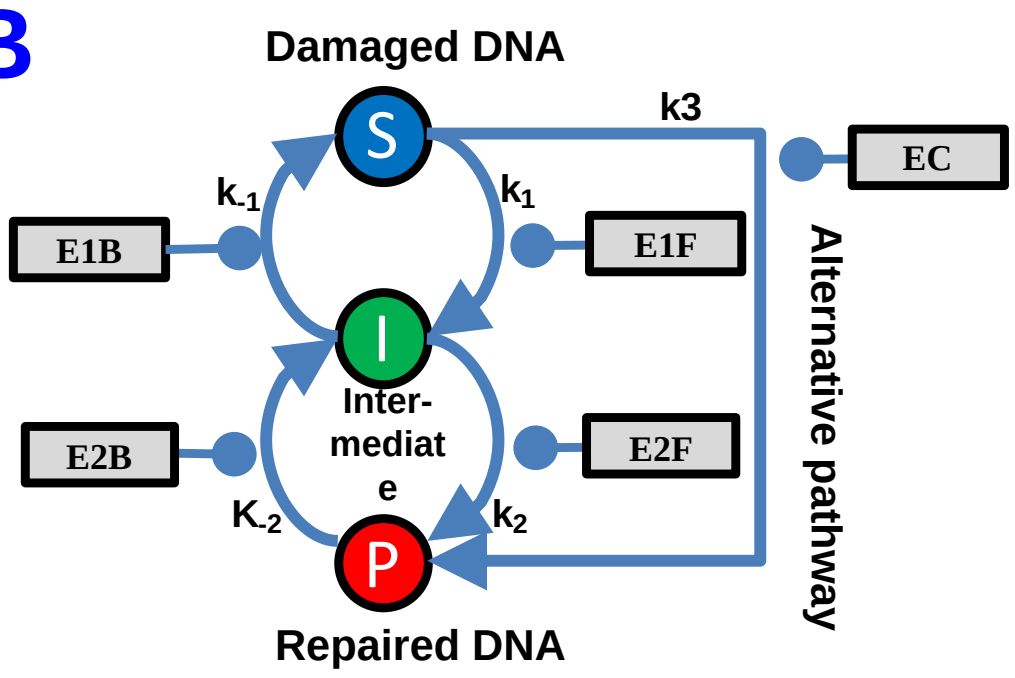

**C**

Pathway steady states for various combinations of parameters in the model

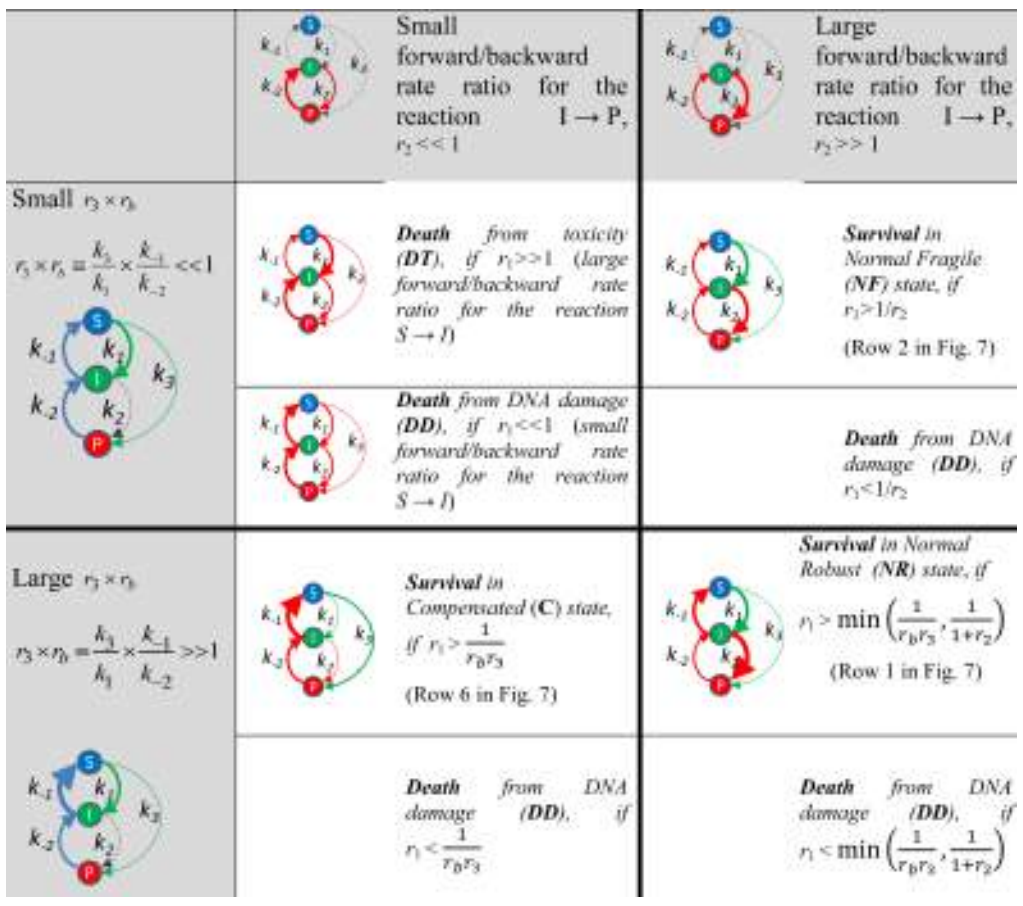

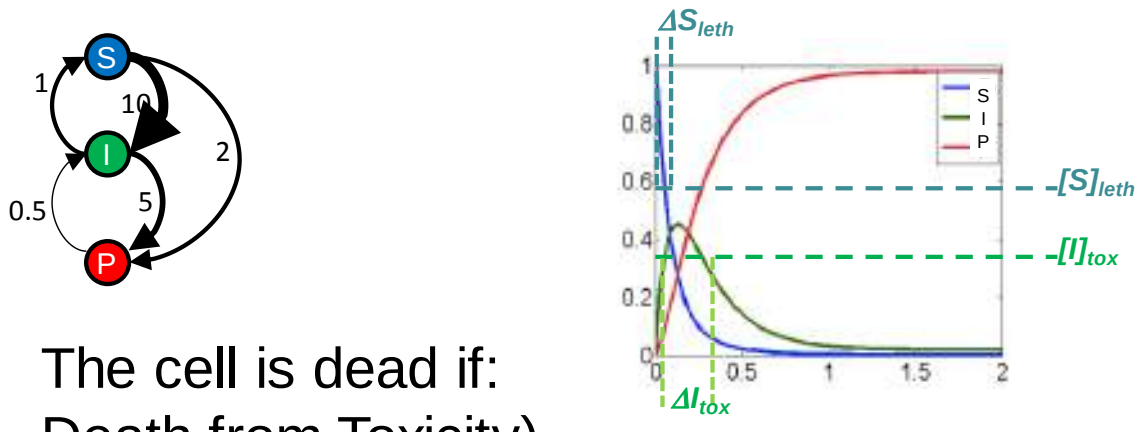

The cell is dead if:

Death from Toxicity)

Amount of ***I*** is higher than $[I]_{tox}$ for sufficiently long period of time $\Delta I_{tox}$

Death from DNA Damage)

Amount of ***S*** is higher than $[S]_{leth}$ for sufficiently long period of time $\Delta S_{leth}$

$$\begin{cases} [\dot{S}] = -k_1 \cdot [S] + k_{-1} \cdot [I] - k_3 \cdot [S] \\ [\dot{I}] = k_1 \cdot [S] - k_{-1} \cdot [I] - k_2 \cdot [I] + k_{-2} \cdot [P] \\ [\dot{P}] = k_2 \cdot [I] - k_{-2} \cdot [P] + k_3 \cdot [S] \end{cases}$$

**D**

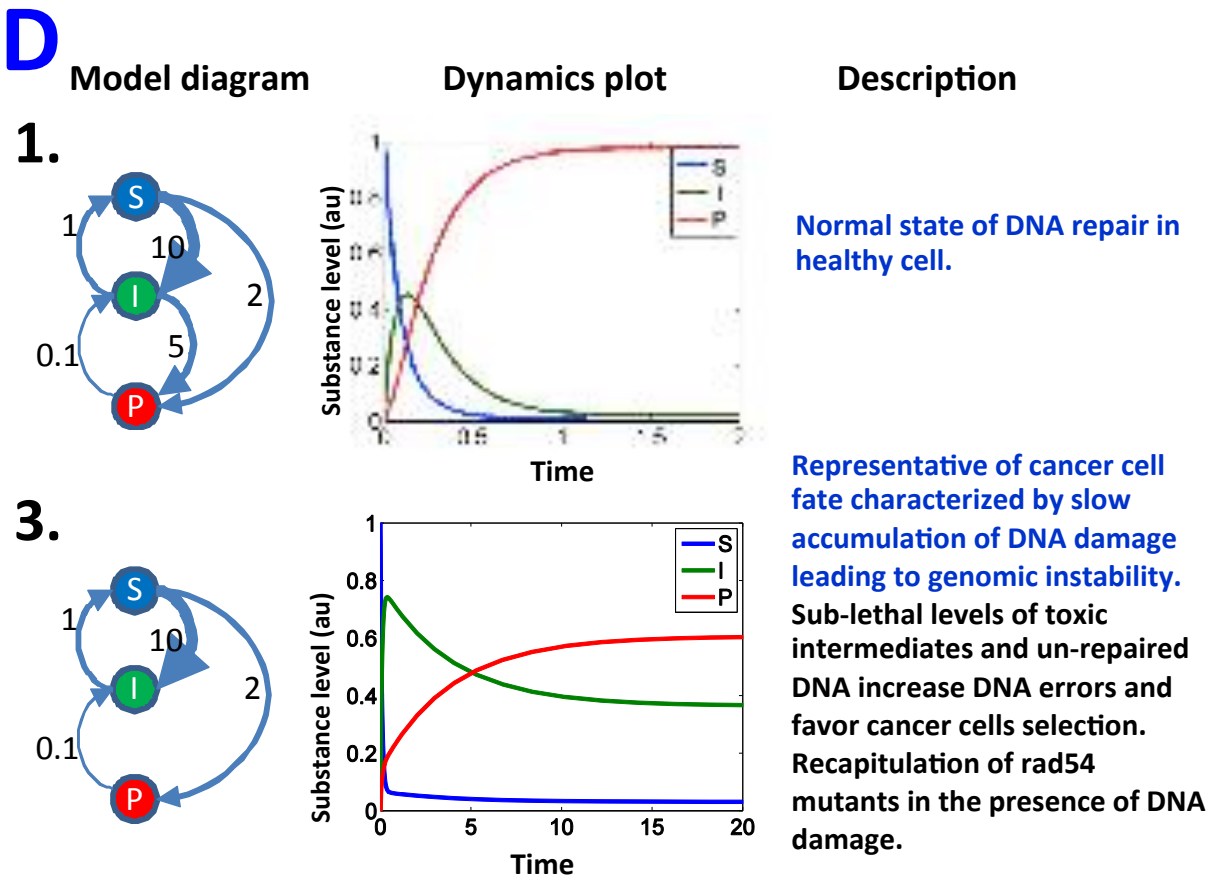

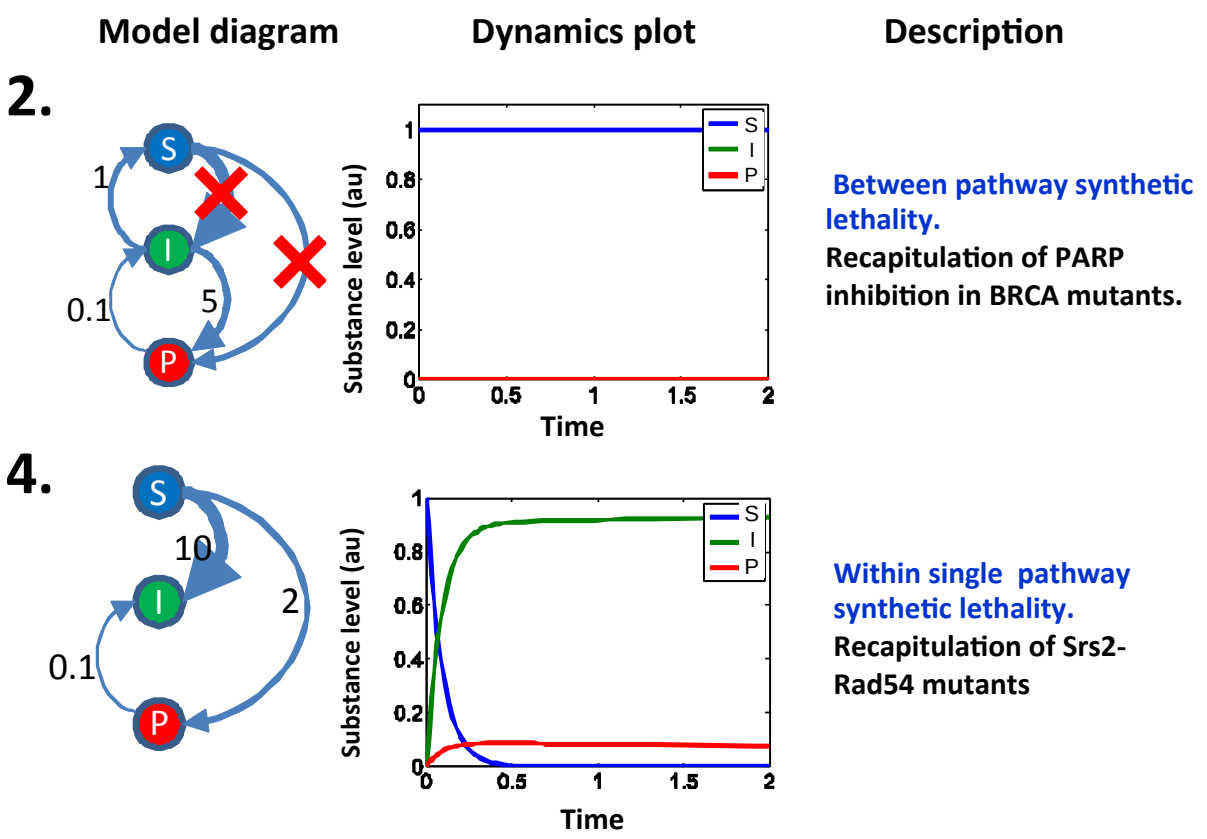

**E**

Normalized proportion of synthetic lethal interactions within one pathway (KEGG pathway database)

| # | KEGG Pathway | Pathway size | Number of negative interactions within pathway | Number of negative interactions, not in complexes | Normalized proportion of negative interactions |
|---|---|---|---|---|---|
| 1 | ***Homologous recombination*** | ***20*** | ***27*** | ***26*** | ***0.137*** |
| 2 | N-Glycan biosynthesis | 31 | 59 | 46 | 0.099 |
| 3 | Various types of N-glycan biosynthesis | 30 | 49 | 35 | 0.080 |
| 4 | ***Non-homologous end-joining*** | ***10*** | ***3*** | ***3*** | ***0.067*** |
| 5 | Biosynthesis of unsaturated fatty acids | 9 | 2 | 2 | 0.056 |
| 6 | Steroid biosynthesis | 14 | 4 | 4 | 0.044 |
| 7 | ***Mismatch repair*** | ***20*** | ***8*** | ***7*** | ***0.037*** |
| 8 | Nicotinate and nicotinamide metabolism | 14 | 3 | 3 | 0.033 |
| 9 | Glycerophospholipid metabolism | 34 | 17 | 17 | 0.030 |
| 10 | Terpenoid backbone biosynthesis | 18 | 4 | 4 | 0.026 |
| 11 | Sphingolipid metabolism | 13 | 2 | 2 | 0.026 |
| 12 | Cell cycle - yeast | 125 | 196 | 176 | 0.023 |
| 13 | ***Base excision repair*** | ***17*** | ***4*** | ***3*** | ***0.022*** |
| 14 | DNA replication | 30 | 10 | 9 | 0.021 |
| 15 | Valine, leucine and isoleucine biosynthesis | 11 | 1 | 1 | 0.018 |
| 16 | Lysine biosynthesis | 11 | 1 | 1 | 0.018 |
| 17 | Phagosome | 34 | 12 | 10 | 0.018 |
| 18 | Protein processing in endoplasmic reticulum | 80 | 69 | 54 | 0.017 |
| 19 | Citrate cycle (TCA cycle) | 33 | 9 | 9 | 0.017 |
| 20 | RNA degradation | 58 | 38 | 28 | 0.017 |
| 21 | Pyruvate metabolism | 33 | 8 | 8 | 0.015 |
| 22 | Propanoate metabolism | 12 | 1 | 1 | 0.015 |
| 23 | MAPK signaling pathway - yeast | 57 | 25 | 23 | 0.014 |
| 24 | Nitrogen metabolism | 22 | 3 | 3 | 0.013 |
| 25 | Protein export | 22 | 5 | 3 | 0.013 |
| 26 | Other types of O-glycan biosynthesis | 13 | 2 | 1 | 0.013 |
| 27 | Meiosis - yeast | 127 | 103 | 87 | 0.011 |
| 28 | Pentose phosphate pathway | 28 | 4 | 4 | 0.011 |
| 29 | Glycosylphosphatidylinositol(GPI)-anchor biosynthesis | 25 | 3 | 3 | 0.010 |
| 30 | Pantothenate and CoA biosynthesis | 15 | 1 | 1 | 0.010 |

**F**

Kinetic trap is a generic feature of cell pathways

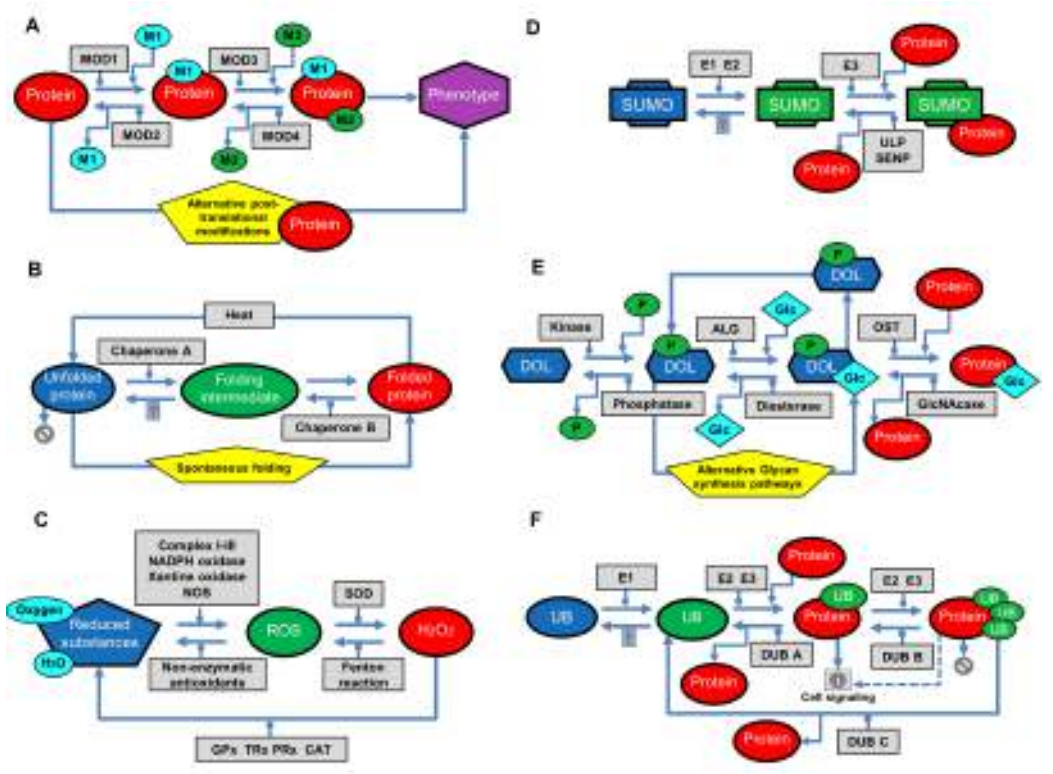

*Figure 10. **Kinetic trap model of synthetic lethality within a single reversible pathway**. (A). DNA repair pathway with reversible step. (B). Abstract representation of Homologous recombination (HR) pathway with reversible steps, containing substrate (S), intermediate (I), product (P) and alternative, compensatory pathway. (C). Pathway steady states for various combinations of parameters in the mathematical model of DNA repair with reversible steps and a toxic intermediate. (D). Modeling possible scenarios of single and double synthetic lethal mutations in the simplest model. The edge thickness denotes the relative speeds between different states of DNA. Scenario '4' represents within singe pathway kinetic trap model of synthetic lethality. (E). Normalized proportion of synthetic lethal interactions within single pathway across KEGG pathway database, calculated using data on the genome-wide screening of genetic interactions in yeast* (Costanzo *et al.*, 2010). *(F). Examples of molecular processes with alternative pathways and potential to kinetic trap into a toxic intermediate.*

These results significantly broaden the interpretation of synthetic lethal effects, which fundamentally impacts on understanding of pathways propagation in the cell. The concept of synthetic lethality has been applied to cancer therapy, and our modeling results suggest new cancer therapy, targeting a single pathway to induce synthetic lethality in pathological cells by trapping reversible pathways into the step where toxic intermediates will accumulate in cancer cells (Zinovyev *et al.*, 2013).

## 4.2 Synthetic lethality mechanisms and organizational principles of cell signaling at different scales

***Working hypothesis***

Availability of high-throughput techniques makes it possible to assay genetic interactions of many genes in parallel in systematic way (high-throughput screens). This data on genetic interactions together with the latest findings on signaling and metabolic pathways helps building higher level models of cell organization and discovering new mechanisms of synthetic interactions. Despite systematic phenomenological classification of genetic interactions, we still lack an exhaustive classification of the mechanistic principles of the extreme case of negative genetic interaction, synthetic lethality. We describe mechanisms of synthetic lethality at several scales, starting from molecular complexes, through molecular pathways and up to the level of functional modules and cells.

***Methodologies and results***

Synthetic lethal interactions are frequently observed within or between members of molecular complexes. Absolute loss of catalytic activity in the essential complex may result in cell death (Figure 11A). It was estimated that among all synthetic lethal genetic interaction pairs, 9-14% belong to the same biological pathway. The mechanisms explaining synthetic lethal interactions in the same pathway vary depending on the architecture of the pathway. There are synthetic lethal mechanisms as defect accumulation, internal redundancy, intrinsic buffering and kinetic trap (Figure 11B).

The most known interpretation of synthetic lethality is when two genes function in parallel pathways that collectively contribute to a common essential biological function. However, in a view of recent findings, this classical paradigm of between pathways synthetic lethality was extended and can be subcategorized in three classes of synthetic lethality mechanisms as redundancy, buffering and synergy (Figure 11C).

The phenotype readout used in the large-scale genetic screenings is the cell growth, which is complex phenotype with many cellular functions contributing. Cell growth is a combination of cell death, cell replication and cell senescence rates, etc. and each of these processes can be addressed as an independent phenotype. Moreover, it is possible to study phenotypes not directly connected to cell growth. Clustering of genes by functional similarity led to the observation that synthetic interactions are more frequent between functional modules with related biological functions rather than between functional modules that have independent cellular roles. For example, genes involved in the functional module of cell polarity more frequently have synthetic interactions with genes involved into the cytoskeleton remodeling module. Genes related to the checkpoints module synthetically interact with the genes from the sister-chromatid cohesion, DNA replication and DNA repair module. Genes responsible for protein folding synthetically interact with the genes from Endoplasnatic reticulum (ER) module, etc. Therefore, it is more probable to anticipate synthetic lethal combinations in functionally-related modules (Figure 11D, upper panel).

In cancer biology, many phenotypes contributing to the disease, are dictated by different cells. For example, phenotypes as Epithelial-Mesenchymal Transition (EMT), different modalities of cell death (apoptosis, necrosis), types of DNA repair mechanisms (single strand, double strand, stalled replication fork) are mostly related to cancer cell. Whereas immune response (innate, adaptive), vascularization signaling (angiogenesis), etc., are dictated by different types of cells. Synthetic interactions at the level of different cell type should be considered. Systematic description of orchestration between these players will allow to point to multi-phenotypic synthetic lethality (Figure11D, lower panel). (Kuperstein *et al.*, in preparation).

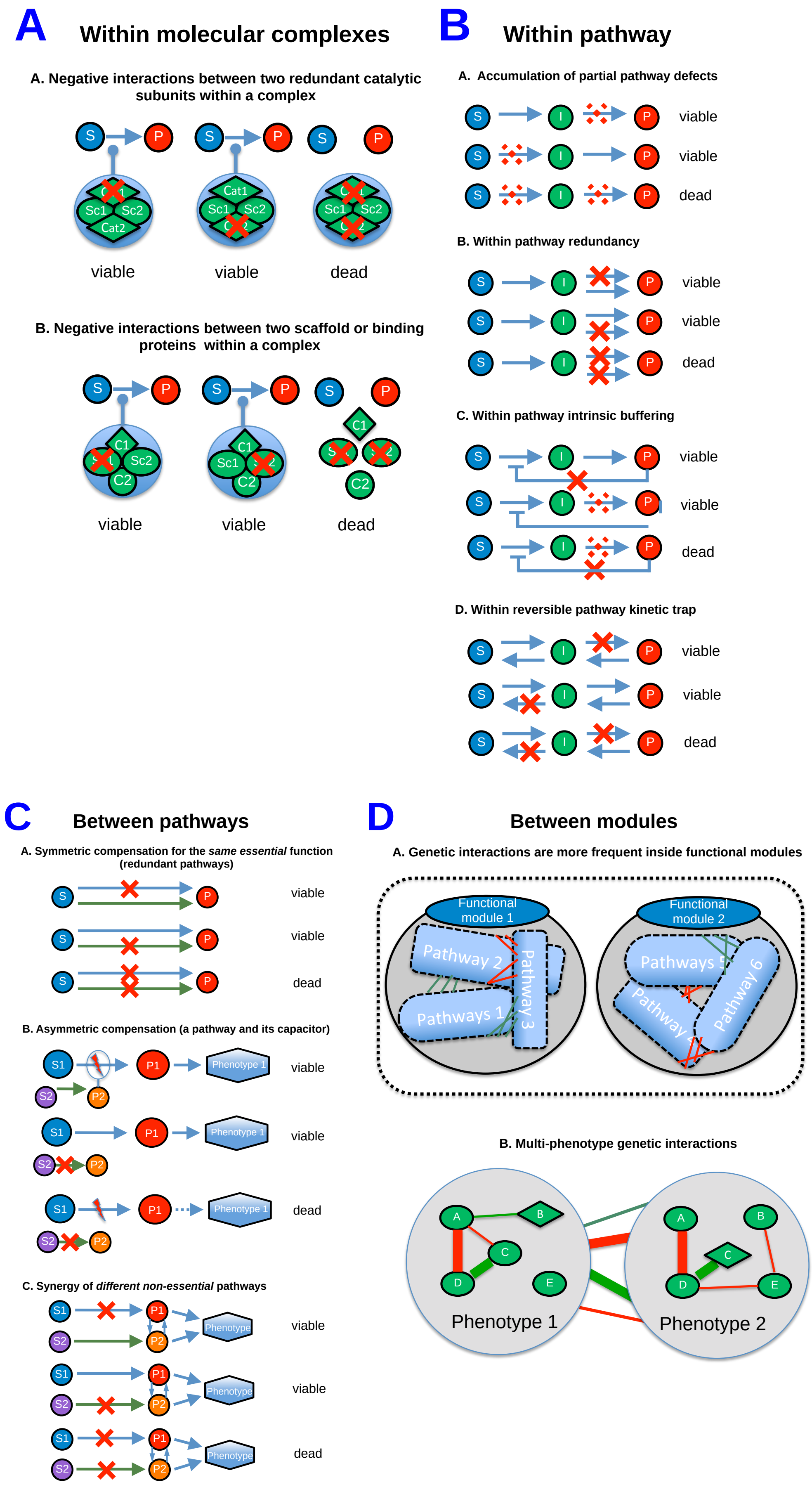

*Figure 11. Mechanisms of synthetic lethality as different scales of cell signaling.*

# CONCLUSIONS

A fundamental question of signaling regulation in human disorders can be addressed by experimental and computational approaches, together helping to understand the principles of signaling rewiring. It will have an impact on personalized intervention schemes, in particular those based on combination of drugs.

Organization and formal representation of the knowledge of cell signaling followed by analysis of network features will provide a more global view on molecular mechanisms and will facilitate modeling of cell fates. It has a potential to explain mutant phenotypes, and reveal new molecular interactions. Considering the topology of signaling networks and studying network perturbations together with high-throughput data can guide us toward an optimal intervention strategy in patients or in studied experimental systems.

Current advances in systems biology and systems medicine allow a standard, unambiguous representation of molecular processes involved in different human diseases, providing a comprehensive platform to interpret experimental results. The collective research effort on different human disorders will lead to the identification of emerging disease hallmarks (Mazein *et al.*, in preparation).

**Typical systems biology project workflow**

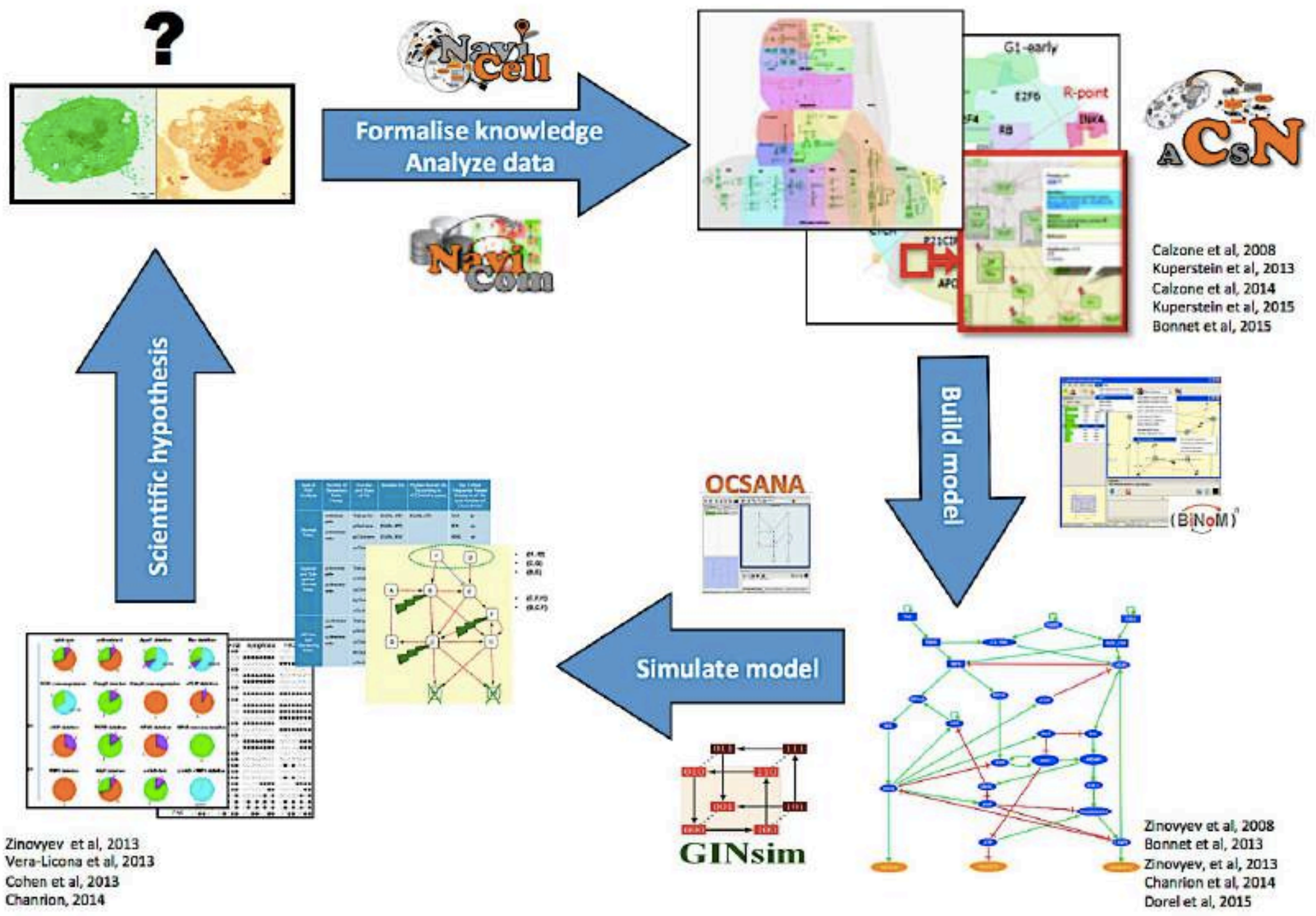

# REFERENCES

**Note 1:** Bolded are papers with my contribution
**Note 2:** F. and I. Kuperstein is the same name

# APPENDIX

## Complete list of publications and communications

All publications are available at: https://www.researchgate.net/profile/Inna_Kuperstein/publications

**Note 2:** F. and I. Kuperstein is the same name

**2015**

1. Russo C et al., Complex intervention sets derived from data-driven network analysis for cancer patients resistant to genotoxic treatment ***(in preparation***)

2. **Kuperstein I** et al., Mechanisms of synthetic lethality at multiple levels of cellular processes ***(in preparation***)

3. Mazein, A. et al., Systems medicine disease maps: prior knowledge is power ***(in preparation)***

4. Jdey W et al., Dbait and PARP inhibitors: Birds of different feathers can also flock together ***(in preparation***)

5. Dorel M, Viara E, Barillot E, Zinovyev A* and **Kuperstein I***. NaviCom: Python package and web interface to create interactive molecular network portraits using multi-level omics data ***(in revision***)

6. Kondratova M, Barillot E, Zinovyev A, **Kuperstein I**. From signaling map construction to data visualization and interpretation (***invited paper in the Bio-Molecular Modeling book, Austin Publishing Group; in press***)

7. Dorel M, Barillot E, Zinovyev A and **Kuperstein I.** Network-based approaches for drug response prediction and targeted therapy development in cancer ***Biochem Biophys Res Commun.*** *doi: 10.1016/j.bbrc.2015.06.094 (2015)* PMID: 26086105

8. **Kuperstein I**, Bonnet E, Nguyen HA**,** Grieco L, Cohen D, Viara E, Kondratova M, Fourquet S, Calzone L, Russo C, Dutreix M, Barillot E and Zinovyev A. Atlas of Cancer signaling Network: navigating cancer biology with Google Maps ***Oncogenesis.*** *doi: 10.1016/j.bbrc.2015.06.094 (2015)* PMID: 26192618

9. **Kuperstein I**, Robine S and Zinovyev A. Network biology elucidates metastatic colon cancer mechanisms. ***Cell cycle*** *doi: 10.1080/15384101.2015.1060816 (2015)* PMID: 26083805

10. Bonnet E, Viara E, **Kuperstein I**, Calzone L, Cohen D., Barillot E and Zinovyev A. NaviCell Web Service for network-based data visualization. ***Nucleic Acids Res***. doi:10.1093/nar/gkv450 (2015) PMID: 25958393

11. **Kuperstein I**, Grieco L, Cohen D, Thieffry D, Zinovyev A and Barillot E. The shortest path is not the one you know: application of biological network resources in precision oncology research. ***Mutagenesis***, doi: 10.1093/mutage/geu078 (2015) PMID: 25688112

**2014**

12. Chanrion M*, **Kuperstein I***, Barrière C, El Marjou F, Cohen D, Vignjevic D, Stimmer L, Dos Reis Tavares S, Cacheux W, Meseure D, Fre S, Martignetti L, Paul-Gilloteaux P, Fetler L, Barillot E, Louvard D, Zinovyev A and Robine S. Notch activation and p53 deletion induce EMT-like processes and metastasis in a novel mouse model of intestinal cancer. ***Nature Communications*** 5:5005. doi: 10.1038/ncomms6005 (2014) PMID: 25295490

13. Calzone L*, **Kuperstein I***, Cohen D, Grieco L, Bonnet E, Servant N, Hupé P, Zinovyev A, Barillot A. Biological network modeling and precision medicine in oncology. ***Bulletin du Cancer*** 101 (1): S18- 21 (2014) PMID: 24966078

**2013**

14. **Kuperstein I***, Cohen D*, Pook S*, Viara E, Calzone L, Barillot E, Zinovyev A. NaviCell: a web-based environment for navigation, curation and maintenance of large molecular interaction maps. ***BMC Syst Biol***., 7:100 doi: 10.1186/1752-0509-7-100. (2013) PMID: 24099179

15. Cohen D, **Kuperstein I**, Barillot E, Zinovyev A, Calzone L. From a biological hypothesis to the construction of a mathematical model. ***Methods Mol Biol.*** 1021:107-25 doi: 10.1007/978-1-62703-450-0_6. (2013) PMID: 23715982

16. Zinovyev A*, **Kuperstein I***, Barillot E, Heyer WD. Synthetic lethality between gene defects affecting a single non-essential molecular pathway with reversible steps. ***PLoS Comput Biol***. 29(4):e1003016. doi: 10.1371/journal.pcbi.1003016. (2013) PMID: 23592964

**2001-2010**

17. **Kuperstein I,** Broersen K, Benilova I, Rozenski J, Jonckheere W, Debulpaep M, Vandersteen A, Segers-Nolten I, Van Der Werf K, Subramaniam V, Braeken D, Callewaert G, Bartic C, D'Hooge R, Martins IC, Rousseau F, Schymkowitz J, De Strooper B. Neurotoxicity of Alzheimer's disease Aβ peptides is induced by small changes in the Aβ42 to Aβ40 ratio. ***EMBO J.*** 29(19):3408-3420 (2010) PMID: 20818335

18. Martins IC*, **Kuperstein I***, Wilkinson H, Maes E, Vanbrabant M, Jonckheere W, Van Gelder P, Hartmann D, D'Hooge R, De Strooper B, Schymkowitz J, Rousseau F. Lipids revert inert Abeta amyloid fibrils to neurotoxic protofibrils that affect learning in mice. ***EMBO J***. 27:224-233 (2008) PMID: 18059472

19. **Kuperstein F**, Eilam R, Yavin E. Altered expression of key dopaminergic regulatory proteins in the postnatal brain following perinatal n-3 fatty acid dietary deficiency. **J Neurochem**. 106(2):662-671 (2008) PMID: 18410511

20. **Kuperstein F**, Yakubov E, Dinerman P, Gil S, Eylam R, Salem N Jr, Yavin E. Overexpression of dopamine receptor genes and their products in the postnatal rat brain following maternal n-3 fatty acid dietary deficiency. **J Neurochem**. 95(6):1550-1562 (2005) PMID: 16305626

21. Yakubov E, Dinerman P, **Kuperstein F**, Saban S, Yavin E. Improved representation of gene markers on microarray by PCR-Select subtracted cDNA targets. ***Brain Res Mol Brain Res*** 137(1-2):110-118 (2005) PMID: 15950768

22. **Kuperstein F**, Brand A, Yavin E. Amyloid Abeta1-40 preconditions non-apoptotic signals in vivo and protects fetal rat brain from intrauterine ischemic stress. ***J Neurochem***. 91(4):965-974 (2004) PMID: 15525350

23. **Kuperstein F**, Yavin E. Pro-apoptotic signaling in neuronal cells following iron and amyloid beta peptide neurotoxicity. ***J Neurochem***. ;86(1):114-125 (2003) PMID: 12807431

24. **Kuperstein F**, Yavin E. ERK activation and nuclear translocation in amyloid-beta peptide- and iron-stressed neuronal cell cultures. ***Eur J Neurosci.*** 16(1):44-54 (2002) PMID: 12153530

25. **Kuperstein F**, Reiss N, Koudinova N, Yavin E. Biphasic modulation of protein kinase C and enhanced cell toxicity by amyloid beta peptide and anoxia in neuronal cultures. ***J Neurochem.*** 76(3):758-767 (2001) PMID: 11158247

# Proceeding papers

1. **Kuperstein I**, Zinovyev A, Barillot E, Heyer WD. Mathematical Modeling of Synthetic Lethality With Applications to DNA Repair. **European Journal of Cancer** 48, S151(2012) Link

2. **Kuperstein** I, Zinovyev A, Cohen D, Fourquet S, Calzone L, Pook S, Vera-Licona P, Bonnet E, Rovera D, Barillot E. Towards of Atlas of Cancer Signaling Networks–Basis for the Institut Curie Systems Biology Platform for Data Analysis and Interpretation. ***European Journal of Cancer*** 48, S152 (2012) Link

3. Vera-Licona P, Zinovyev A, Bonnet E, **Kuperstein I**, Kel-Margoulis O, Kel A, Dubois T, Tucker G, Barillot E A Pathway-based Design of Rational Combination Therapies for Cancer, ***European Journal of Cancer*** 48, S154 (2012) Link

4. Vera-Licona P, Zinovyev A, Bonnet E, **Kuperstein I**, Kel-Margoulis O, Kel A, Dubois T, Tucker G,

Barillot E. A signaling pathway rationale for the design of combination therapies for cancer. **Cancer Research**, 72 (13): A1 (2012) Link

5. **Kuperstein I**, Vera-Licona P, Zinovyev A, Tucker GC, Dubois T, Barillot E. Integrated cell cycle and DNA repair signaling network modeling for identification of key molecular regulators in basal-like breast cancer. **European Journal of Cancer**, Supplements 8 (5), 208-209 (2010) Link

6. Benilova I, Chong SA, **Kuperstein I**, Broersen K, Schymkowitz J, Rousseau F, Bartic C, Callewaert G, De Strooper B. The A-beta 42/40 ratio is a driver of acute synaptotoxicity and LTP impairment. **Proceedings of the 7th Meeting on Substrate-Integrated Microelectrode Arrays**. pp.159-160 (2010) Link

7. Benilova I, **Kuperstein I**, Broersen K, Schymkowitz J, Rousseau F, Bartic C, De Strooper B. MEA Neurosensor, the Tool for Synaptic Activity Detection: Acute Amyloid-β Oligomers Synaptotoxicity Study. **Proceedigs of World Congress on Medical Physics and Biomedical Engineering**, pp. 314-316 (2010) Link

8. Brouillette J, **Kuperstein I**, Ahmed T, Van der Jeugd A, Burnouf S, Belarbi K, Blum D, Balschun D, D'Hooge R, De Strooper B, Buee L. Tau experimental models: effects of amyloid beta oligomers on tau pathology. **Proceedings of the 9th International Conference AD/PD**, pp. 2057 (2009)

9. **Kuperstein I**, Martins I, Wilkinson H, Vanbrabant M, Maes E, D'Hooge R, Schymkowitz J, Rousseau F, De Strooper B. Lipids revert inert amyloid fibrils to neurotoxic protofibrils that affect learning in mice. **Alzheimer's & Dementia** 4(4):T134 (2008) Link

10. Yavin E, Yakubov E, Dinerman P, Gil S, Brand A, **Kuperstein F**. Adequate or deficient omega-3 diets and the perinatal brain. **J Neurochem**. 90, 82-82 (2004)

11. **Kuperstein F** and Yavin E. Role of Fe2+ in Abeta1-40 induced signal transduction and cell cytotoxicity in cerebral neuron cultures. **Biochemical society transactions** 28 (5), A431-A431 (2000)

12. Yavin E, Reiss N, Gil S, **Kuperstein F**, Glozman S. Lipid signal molecules and their regulation during oxidative stress in the developing brain **J Neurochem.** 72, S90-S90 (1999)

13. **Kuperstein F**, Koudinova N, Yavin E. Regulation of PKC and MAPK-related phosphorylation cascades by beta-amyloid (1-40) during oxygen deprivation in rat cerebral neurons **Neuroscience letters**, S24-S24 (1998)

14. **Kuperstein F**, Seger R, Yavin E. Stress-induced activation of PKC and MAPK isoforms in transformed oligodendroglia cells after heat shock **J Neurochem**. 70, S14-S14 (1998)

15. **Kuperstein F**, Bercovici E, Seger R, Yavin E. Protein phosphorylation is modulated in fetal brain preparations after episodes of ischemia **Neuroscience Letters** 237, S30 (1997)

# Invited talks

**2015**

1. The Systems Biology Institute, Tokyo, Japan (Invited by H. Kitano)
Talk title: Atlas of Cancer Signaling Network and NaviCell: comprehensive resource and web tool for data analysis and interpretation
2. Kyoto University Bioinformatics Seminar Series, Kyoto, Japan (Invited by M. Kanehisa)
Talk title: Atlas of Cancer Signaling Network and NaviCell: comprehensive resource and web tool for data analysis and interpretation

**2013**

3. INCA NATIONAL MEETING PROGRAM "RADIOBIOLOGY IN MEDICINE", Paris, France (Invited by M. Dutreix)
Talk title: Comprehensive map of DNA repair signalling network: computational analysis of synthetic lethality in DNA repair and application for cancer treatment design
4. Advanced Topics in Genomics and Cell Biology workshop. Unicamp, Campinas, Sao Paulo, Brazil (Invited by E. L. Sartorato)
Talk title: Synthetic lethality mechanisms for cancer treatment: computational analysis of synthetic lethality in DNA repair pathway and beyond
5. AC Camargo Hospital, Sao Paulo, Brazil Invited seminar and Post-graduate course (Invited by S. Rogatto)

Talk title: Comprehensive map of DNA repair signaling network: computational analysis of synthetic lethality in DNA repair and application to cancer treatment.

6. VIB Departmental seminar on Human Genetics, LU Leuven/VIB, Leuven, Belgium (Invited by B. De Strooper)
   Talk title: A systems biology approach for clinical applications and basic research:signalling networks construction, analysis and modeling.

**2011**

7. Cancer Systems Biology Centre, NKI-AVL, Amsterdam, Netherlands (Invited by L. Wessels)
   Talk title: Curie Atlas of Cancer Signaling Networks: a systems biology approach for basic research and clinical applications
8. Departmental seminar on Microbiology and Molecular Genetics, University of California, Davis, US (Invited W.D. Heyer)
   Talk title: Signalling networks construction and modelling: a systems biology approach for basic research and clinical applications

# Conference presentations (from 2005 to present)

**2015**

1. International conference on systems Biology (ICSB 15), Singapore, November 2015 (two poster)
2. ICSB workshop-BioNetVisA: From biological network reconstruction to data visualization and analysis in molecular biology and medicine, Singapore, November 2015 (**oral presentation**)
   Talk title: Prediction of sensitivity to genotoxic drug by modeling cancer cell lines and patient omics data in the context of comprehensive DNA repair signaling network
3. 2nd CRCL International Symposium on: "The Tumor and its Microenvironment: Challenges and Innovative Therapies" (CRCL), Lyon, France , September 2015 (poster)
4. GIW/InCoB, Tokyo, Japan, September 2015 (**two oral presentations**)
   Talk 1 title: Atlas of Cancer Signaling Network: a systems biology resource for integrative analysis of cancer data with Google Maps
   Talk 2 title: Finding metastasis inducers in colon cancer through network analysis: concomitant Notch activation and p53 deletion trigger the process
5. Special interest group meeting - Networks Biology (SIG-NetBio), ISMB/ECCB Dublin, Ireland, 2015 (**oral presentation**; poster)
   Talk title: Data visualization and modeling using Atlas of Cancer Signaling Network predicts clinical outcome
6. International conference on Intelligent Systems for Molecular Biology and European conference of computational biology (ISMB/ECCB), Dublin, Ireland, 2015 (**highlight paper oral presentation**, **late breaking research oral presentation**, poster)
   HR talk title: Finding metastasis inducers in colon cancer through network analysis: concomitant Notch activation and p53 deletion trigger the process
   LBR talk title: Data visualization and modeling using Atlas of Cancer Signaling Network predicts clinical outcome
7. cBio: 2nd Symposium on Complex Biodynamics & Networks, Tsuruoka, Japan, 2015 (**oral presentation**, poster)
   Talk title: Modeling a comprehensive DNA repair signaling network from ACSN collection predicts genotoxic drug sensitivity from cancer cell lines and patient data

**2014**

8. 15th International Conference on Systems Biology (ICSB14), Melbourne Australia, 2014 (**oral presentation**, poster)
   Talk title: Visualization and analysis of data using Atlas of Cancer Signalling Networks (ACSN) and NaviCell tools for integrative systems biology of cancer
9. International conference on Intelligent Systems for Molecular Biology (ISMB), Boston, USA (**highlight paper oral presentation** - replacement presenter E. Barillot)
   Talk title: NaviCell: a web-based environment for navigation, curation, maintenance and data analysis in the context of large molecular interaction maps
10. ECCB14, Strasbourg, France, 2014 (poster)
11. ECCB 14workshop-BioNetVisA: From biological network reconstruction to data visualization and analysis in molecular biology and medicine, Strasbourg, France, 2014 (**oral presentation**, poster)
    Talk title: From biological network reconstruction to data visualization and analysis in molecular biology and medicine
12. Biocuration2014, Toronto, Canada, 2014 (**oral presentation**, poster)
    Talk title: Visualization and analysis of data using Atlas of Cancer Signalling Networks (ACSN) and NaviCell tools for integrative systems biology of cancer

**2013**

13. EMBO conference: 'The DNA damage response in cell physiology and disease', Cape Sounio, Greece, 2013 (poster)
14. The Computational Modeling in Biology Network (COMBINE2013), Paris, France, 2013 (**oral presentation**)
    Talk title: Atlas of Cancer Signaling Networks (ACSN) and NaviCell are user-friendly web-based environments for integrative systems biology of cancer
15. Moscow Conference on Computational Molecular Biology (MCCMB'13), Moscow, Russia, 2013 (**oral presentation**)
    Talk title: Modeling signaling networks for explaining synthetic gene interactions leading to invasive phenotype

in colon cancer mouse model

16. International conference on Intelligent Systems for Molecular Biology and European conference of computational biology (ISMB/ECCB), Berlin, Germany, 2013 (**highlight paper oral presentation**, two posters)
    Talk title: Synthetic lethality between gene defects affecting a single non-essential molecular pathway with reversible steps
17. Special interest group meeting - Networks Biology (SIG-NetBio), ISMB/ECCB Berlin, Germany, 2013 (**oral presentation; network construction panel presenter**; poster)
    Talk title: Atlas of Cancer Signaling Networks (ACSN) and NaviCell are user-friendly web-based environments for integrative systems biology of cancer
18. NUS-ENS workshops–Novel genome-wide approaches to decipher transcriptional and epigenetic regulation in mammalian cells, Paris, France, 2013

**2012**

19. Biocuration2013, Cambridge, UK, 2013 (two posters)
20. Colloque - Bioinformatique et cancer, Paris, France, 2012
21. European Association of Cancer Research Congress (EACR2012), Barcelona, Spain, 2012 (two posters)
22. Integrative Network Biology 2012: Network Medicine, Helsingør, Denmark, 2012 (two posters)

**2011**

23. EPIGENETICS: FROM BASES TO PATHOLOGY, Paris, France, 2011
24. International Society for Computational Biology and European Conference on Computational Biology meeting(ISCB/ECCB), Vienna, Austria, 2011 (poster)
25. Cancer Proteomics2011, Dublin, Ireland, 2011 (poster)
26. Triple negative breast cancer conference, London, UK, 2011 (poster)
27. Breast cancer conference, Madrid, Spain, 2011 (**oral presentation**)
    Talk title: Integrated cell cycle and DNA repair signaling network modeling for identification of key molecular regulators in basal-like breast cancer

**2010**

28. EpiGeneSys Kick-Off Meeting, Paris, France, 2010
29. The 10th International Conference on Systems Biology (ICSB2010), Edinburgh, Scotland, 2010 (poster)
30. European Association for Cancer Research meeting (EACR2010), Oslo, Norway, 2010 (poster)
31. Second Annual RECOMB Satellite Workshop on Computational Cancer Biology, Oslo, Norway, 2010 (poster)

**2005-2009**

32. Alzheimer’s disease and Parkinson disease meeting (ADPD), Prague, Czech Republic, 2009 (**oral presentation**)
33. MEMOSAD general assembly II, Leuven, Belgium, 2009 (**oral presentation**)
34. International conference of Alzheimer’s disease (ICAD), Chicago, USA, 2008 (**oral presentation**)
35. MEMOSAD general assembly I, Munich, Germany, 2008 (**oral presentation**)
36. IPSEN foundation conferences-Neurodegeneration in Alzheimer’s disease, Paris, France, 2007 (poster)
37. Protein misfolding and aggregation in ageing and disease, Roscoff, France, 2007 (**oral presentation**)
38. Alzheimer’s disease and Parkinson disease meeting (ADPD), Salzburg, Austria, 2007 (poster)
39. NeuroNe workshop-Dysfunction of axons and synapses in neurodegeneration, Cambrige, UK, 2006 (poster)